\documentclass[fleqn,usenatbib]{mnras}

\usepackage[T1]{fontenc}
\usepackage{ae,aecompl}
\usepackage{graphicx}
\usepackage{longtable}
\usepackage[varg]{txfonts}
\usepackage{color}
\usepackage{epsfig}
\usepackage{threeparttable}
\usepackage{threeparttablex}
\usepackage{url}
\usepackage{textcomp}
\usepackage{multirow}
\title[Polarization models of stars with debris disks]{Polarization of stars with debris disks: comparing observations with models}

\author[J. Vandeportal et al.]{
Julien Vandeportal,$^{1*}$
\,Pierre Bastien,$^{1}$\thanks{E-mail: bastien@astro.umontreal.ca (PB) vandeportal.julien@gmail.com (JV)}
Am\'elie Simon,$^{1}$
Jean-Charles Augereau,$^{2}$ \newauthor
\'Emilie Storer$^{3}$
\\
$^{1}$D\'epartement de physique and Centre de recherche en astrophysique du Qu\'ebec, Universit\'e de Montr\'eal, C.P. 6128, Succ. centre-ville, \\Montr\'eal, QC H3C 3J7, Canada\\
$^{2}$Univ. Grenoble Alpes, CNRS, IPAG, 38000 Grenoble, France \\
$^{3}$Department of Physics and Centre de recherche en astrophysique du Qu\'ebec, McGill University, 3600 University St., Montr\'eal, QC H3A 2T8, Canada\\
} 

\date{Accepted 2018 October 27. Received 2018 October 12; in original form 2018 July 20}

\pubyear{2018}

\hypersetup{draft}
\begin{document}

\label{firstpage}
\pagerange{\pageref{firstpage}--\pageref{lastpage}}
\maketitle

\begin{abstract}
The $Herschel$ Space telescope carried out an unprecedented survey of nearby stars for debris disks. The dust present in these debris disks scatters and polarizes stellar light in the visible part of the spectrum. We explore what can be learned with aperture polarimetry and detailed radiative transfer modelling about stellar systems with debris disks. We present a polarimetric survey, with measurements from the literature, of candidate stars observed by DEBRIS and DUNES $Herschel$ surveys. We perform a statistical analysis of the polarimetric data with the detection of far-infrared excess by $Herschel$ and $Spitzer$ with a sample of 223 stars. Monte Carlo simulations were performed to determine the effects of various model parameters on the 
polarization level and find the mass required for
detection with current instruments. Eighteen stars were detected with a polarization $0.01 \le P \lesssim 0.1$ per cent and $\ge3\sigma_P$, but only two of them have a debris disk. No statistically significant difference is found between the different groups of stars, with, without, and unknown status for far-infrared excess, and presence of polarization. The simulations show that the integrated polarization is rather small, usually $< 0.01$ per cent for typical masses detected by their far-infrared excess for hot and most warm disks. Masses observed in cold disks can produce polarization levels above $0.01$ per cent since there is usually more dust in them than in closer disks.
We list five factors which can explain the observed low-polarization detection rate. Observations with high-precision polarimeters should lead to additional constraints on models of unresolved debris disks.
\end{abstract}


\section{Introduction}
	
The first large unbiased survey of debris disks was carried out by the Infra-Red Astronomical Satellite ($IRAS$). It was found that $\approx$\,15 per cent of main-sequence stars host debris disks \citep{1993prpl.conf.1253B, 1999A&A...343..496P}. However, this fraction might be as high as 25 per cent since there is evidence for a population of disks too cold to have been detected by $IRAS$ \citep{2003MNRAS.342..876W,2006A&A...460..733L,2007ApJ...660.1556R}. More recently, the DUNES \citep{2013A&A...555A..11E} and DEBRIS \citep{2015A&A...582L...5R} $Herschel$ surveys measured a fraction of stars with disks as high as $20 \pm 2$ per cent  and $17 \pm 2$ per cent, respectively. The full DUNES sample of 177 FGK stars within 20 pc was found to have a fraction of $20^{+7}_{-5}$ per cent \citep{2016A&A...593A..51M} of debris disks. For single and multiple stars, the rates are $21 \pm 3$ per cent and $11 \pm 3$ per cent respectively \citep{2015MNRAS.449.3160R}. \cite{2014A&A...565A..15M} found a fraction of $29 \pm 9$ per cent debris disks associated with planetary systems, combining the DUNES and DEBRIS samples. Such high proportions of stars hosting debris disks indicate that belts of small bodies (asteroids, comets) are common outcomes of the planet formation process, and survive over long timescales. 

Debris disks are found around stars at every age but the lifetime of dust composing them is shorter than that of the hosting star, due to the Poynting--Robertson effect and radiation pressure. However debris disks are still present due to replenishment mechanisms that continuously feed dust to the debris disks \citep{1993prpl.conf.1253B}. These mechanisms are collisions between planetesimals and sublimation of comets \citep{1994Icar..107..117W,2002MNRAS.334..589W,2007A&A...472..169T}. Hence, observing debris disks is an important way to infer the presence of solid bodies around stars and to understand the dynamics of planetary systems \citep{2008ARA&A..46..339W,2010RAA....10..383K}. Moreover, \cite{2005ApJ...622.1160B} found a correlation between the presence of debris disks and the presence of planets. However, it is still a subject of debate \citep{2009ApJ...700L..73K, 2015yCat..35790020M,2014A&A...565A..15M}. In any case, for spatially-resolved disks, their wide extension and the presence of structures indicates that planetesimals and probably even planets are present and perturb the disk \citep{1993prpl.conf.1253B, 2012A&A...544A..61E, 2014AJ....148...59S}.

Observing in the close solar environment provides a sample covering a wide range of stellar parameters for which a rich literature already exists. It also maximizes the possibility of finding disks and spatially resolving them. The space telescope $Herschel$ observed at far-infrared (FIR) and submillimetre (submm) wavelengths in two of its open time key programs DEBRIS and DUNES. The PACS instrument was used at 70/100 and 160~$\mu$m for the surveys, complemented by additional observations with SPIRE at 250, 350 and 500~$\mu$m for some selected stars. The DEBRIS targets are the closest evenly distributed stars along the spectral types A, F, G, K and M and constitute a flux-limited survey. The DUNES survey is volume-limited and includes all FGK stars within 20 pc plus a few more out to 25 pc, observed as deep as necessary to detect the stellar photosphere at 100~$\mu$m. Since the targets were selected only by their distances, DEBRIS and DUNES samples give us statistical information about debris disks: age, stellar mass, metallicity, presence of planets, system morphology, multiplicity, presence of disks, etc\dots Early results from the DEBRIS survey is given by \cite{2010A&A...518L.135M} and an exhaustive review by \cite{2014prpl.conf..521M}. Many stars are in common with the James-Clerk-Maxwell Telescope (JCMT) SCUBA-2 Observations of Nearby Stars (SONS) survey at 450 and 850~$\mu$m (\cite{2010MNRAS.403.1089P,2012yCat..74031089P}, \cite{2007PASP..119..842M}, \cite{2013MNRAS.435.1037P}, \cite{2017MNRAS.470.3606H}). The $Spitzer$ telescope with its IRS and MIPS instruments also provided useful data on the short-wavelength side of the $Herschel$ coverage, between 7 and 70~$\mu$m \citep{2006ApJ...639.1166B,2006ApJ...652.1674B,2013ApJ...768...25G}. Hence the spectral coverage spans from 7 to 850~$\mu$m for these stars. Distances of the farthest stars are 9, 16, 21, 24 and 46 pc for the M, K, G, F and A stars respectively.

The most efficient way to detect disks in surveys is to use the radiation excess in the infrared (IR) and submm compared to the stellar photospheric flux, as the DUNES \citep{2013A&A...555A..11E} and DEBRIS \citep{2015A&A...582L...5R} surveys proceeded. 
This excess radiation comes from thermal emission by dust grains heated by the star. About half of the $Herschel$ disk detections have been resolved \citep{2014prpl.conf..521M}, and 16 resolved debris disks out of 49 detected disks have been observed by the SONS survey \citep{2017MNRAS.470.3606H}. The great advantage of resolved observations is the wealth of information available for modelling the disks.

There are other ways to detect debris disks (e.g., \cite{2010RAA....10..383K}). With adaptive optics and instruments such as GPI and SPHERE one can image directly the dust in the visible and near-infrared (NIR) but this method is time consuming and is not very efficient for large surveys. One measures stellar light scattered by dust in circumstellar disks which are often resolved for nearby stars \citep{2014AJ....148...59S}. \cite{2014prpl.conf..521M} gave an exhaustive list of resolved disks at various wavelengths.

Polarization is an interesting and useful way to perform surveys because stellar photons scattered by dust into our line of sight are polarized. Such polarization has already been measured in spatially resolved disks such as the $\beta$ Pictoris \citep{1991MNRAS.252P..50G,1995Ap&SS.224..395W,2006ApJ...641.1172T} and the AU Microscopii \citep{2007ApJ...654..595G} disks. Others such as \cite{2001A&A...379..564O, 2002Ap.....45...25E, 2005ASPC..343..215T, 2006A&A...452..921C} and \cite{2010AAS...21537403W} presented unresolved polarization measurements of stars with IR excess. \cite{1996A&AS..120..451B} and \cite{2000A&A...362..978B} compared unresolved polarization of stars with circumstellar matter based on their NIR excess and those that are devoid of such matter. This polarization depends on many properties such as the size, shape and composition of the grains. Even if it is unresolved, polarimetry  can also yield the orientation of the disk projected on the plane of the sky, which NIR, FIR or submm excess emission alone cannot provide, unless of course the source is resolved. Two examples of this are given in section \ref{sub:3sigma}, one observed and the other predicted. Polarization is due to scattering, whose components of the electric field perpendicular and parallel to the scattering plane differ, usually such that the polarization is perpendicular to the scattering plane. This plane includes the light emitter, the scatterer and the observer. The optical thickness in debris disks is so low that multiple scattering is negligible. 

In this paper we present a polarimetric survey of 109 stars from the DEBRIS catalogue in section 2. We extend our sample to include measurements in the literature of additional stars from the DUNES \& DEBRIS surveys and carry out a statistical analysis of this larger sample in section 3. 
We present, in section 4, Monte Carlo simulations and use an analytical model to  compare with observations. The discussion, in section 5, considers the effects of interstellar polarization, explains why cold disks have larger polarizations and compares with other recent observations. Finally, in the last section we conclude and offer suggestions for future research.
 	
\section{Observational results}\label{OR}

\subsection{Observations and other data}
	
The observations were made at the 1.6-m Ritchey--Chr\'etien telescope of the Mont-M\'egantic Observatory (OMM), based in Qu\'ebec, Canada. We observed in three runs between 2009~December~1 and 2010~March~3. We used an 8.18\,arcsec aperture, all multiple stars we observed were integrated at the same time in the 8.18\,arcsec aperture. We used a broadband red filter, RG645, which yields a bandpass centered at 766~nm with a FWHM of 241~nm. Polarization was measured with the Beauty and the Beast instrument which is a two-channel photoelectric polarimeter. It uses a Wollaston prism as analyzer, a Pockels cell operated at 125 Hz as modulator, and an achromatic quarter wave plate.

The data were calibrated for polarization efficiency with a prism (between 75 per cent and 83 per cent), for instrumental polarization using unpolarized standards and for the zero point of position angles using polarized standard stars. More information about the instrument and the method of observation is given by \cite{1995PASP..107..483M}. We observed unpolarized standards for each run and used the same ones as PlanetPol \citep{2009MNRAS.393..229L} whenever possible. The polarized standards we observed come from \cite{1990AJ.....99.1243T} and from the PlanetPol list of polarized stars \citep{2006PASP..118.1302H}.

We adjusted the integration time according to the magnitude of the star in order to have an expected polarization uncertainty of $\approx 0.04$ per cent. Uncertainties $\sigma_{P}$ are calculated with photon statistics and also include the previous uncertainty due to calibration mentioned above. The uncertainties range between 0.02 per cent and 0.12 per cent (with a mean of 0.04 per cent). The uncertainty on the polarization position angles (hereafter polarization angles) are computed as usual \citep{SERKOWSKI1962289} with
\begin{equation}
\sigma_{\theta}=28.65\degr \frac{\sigma_P}{P}.
\label{Stheta-eq}
\end{equation}

The data were pre-analysed by the computer `the Beauty' while observing. The rest of the analysis was automated using IDL programs created for this purpose. The data were also corrected for bias in the usual way \citep{SERKOWSKI1962289}. We assumed that circular polarization would be negligible when we did the data reduction. Of the 297 DEBRIS stars that are visible from the observatory 108 were observed for this analysis.  We also observed the star HR\,8799 even if it is not included in the DEBRIS survey, as it is a target of particular interest: it hosts four imaged planets \citep{2008Sci...322.1348M,2010Natur.468.1080M} and debris disks \citep{2009ApJ...705..314S,2009A&A...503..247R}. 

\subsection{Results}
\label{OMM-results}

All targets from the DEBRIS and DUNES surveys are nearby stars with the furthest one at 46 pc. In our analysis we assume that the interstellar polarization is negligible for such distances \citep{1977A&AS...30..213P,1986A&AS...64..487K,1993A&AS..101..551L,2009MNRAS.393..229L}; hence the polarization we measured is intrinsic to the stars (see the discussion in section 5.1 for more information). 

The instrument works in such a way that there is a redundancy in the measurements of the Stokes parameters: measurements were taken at 0\degr, 45\degr, 90\degr~and 135\degr~from a certain reference. Hence measurements at 0\degr~and 90\degr~for example should give approximatively the same results. We compared the data between 0\degr~and 90\degr~and between 45\degr~and 135\degr, for all measurements above 2$\sigma_P$. We found that measurements are coherent, but not strongly correlated. 

\begin{figure}
\centering
\includegraphics[angle=0,origin=bl,width=0.95\columnwidth]{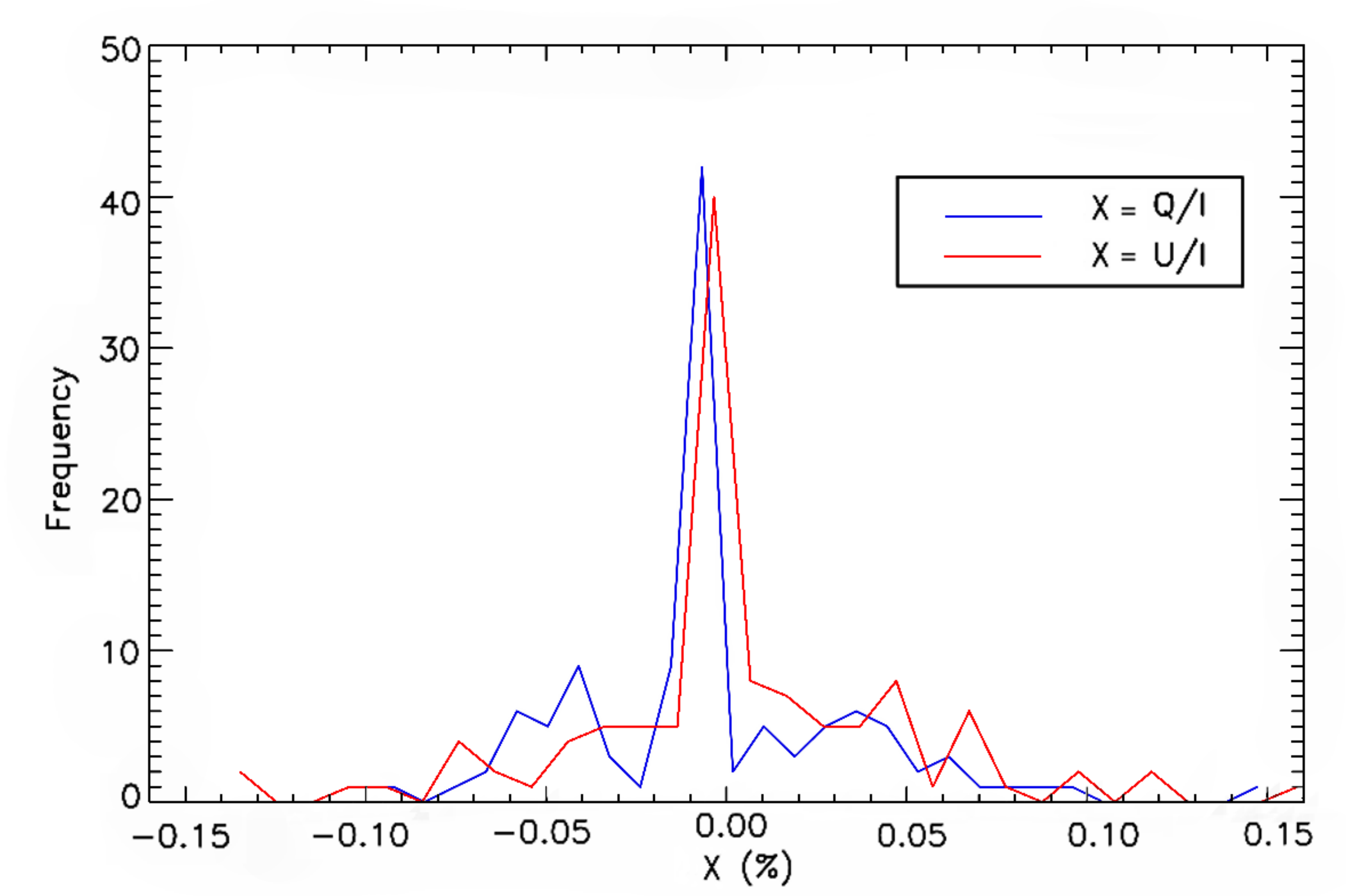}
\caption{Frequency distribution of the normalized Stokes parameters $Q/I$ and $U/I$ for all stars measured at OMM.}
\label{fig1}
\end{figure}

We plot the histograms of stars with a given $Q/I$ and $U/I$ in Figure~\ref{fig1}. Firstly, we can verify that the instrumental polarization is well determined; in this case we should have a peak in the distributions around 0\,per cent. We see that it is the case for $U/I$ but $Q/I$ has a small offset of -0.01\,per cent. It is nonetheless smaller than the uncertainty on the determination of the instrumental polarization, therefore it is compatible with 0\,per cent. A very strong peak around 0\,per cent is seen in both distributions, with two bumps around that peak. This indicates that the uncertainties were larger on certain nights than on others. It is indeed the case as we had some problems with one of the two photomultipliers during some nights and we had to use only half of the data. Finally, we cannot clearly conclude about the presence of polarized stars which would stand in the wings of the peaks. Results are given in Tables~\ref{tab:3sigma} and \ref{tablestars} (Appendix \ref{Appen}) and are presented with the full sample in section \ref{StatComp}.

\subsection{HR 8799}

We observed also HR~8799 even if it is not a target of the DEBRIS survey. We  found a small polarization of 0.07\,per cent at 2.8 times the uncertainty (Table~\ref{tablestars}). This star is particularly interesting since it hosts four resolved planets \citep{2008Sci...322.1348M,2010Natur.468.1080M}. 
\cite{2009ApJ...705..314S} and \cite{2009A&A...503..247R} modelled the IR/submm excess of HR~8799 and found that this star also hosts three debris disks: an inner warm disk, a planetesimal disk and a halo. Confirmation of our 2.8$\sigma$ result and observations in other wavelength bands would provide better constraints on the debris disks.

\subsection{Discussion}

We have made a coherent census of the polarization due to debris disks for 109 stars. We have one detection above 3$\sigma_P$ (HD 115404 = K046) which is what one should expect statistically for such a sample. This rate can be explained by many factors: only $\approx17$ per cent of DEBRIS survey stars were found to have debris disks \citep{2015A&A...582L...5R}; in face-on disks the integrated polarization vector cancels out by symmetry (only a fraction of disks have favourable inclinations for detection) and we have measurement uncertainties of $\approx0.04$ per cent, which seems to be at the limit of detection (see section \ref{Models}). The case of an eccentric disk is considered in the simulations presented below (section \ref{Models}). More importantly, since we used an 8.18\,arcsec aperture centered on the star the unpolarized light from the star is integrated at the same time as the polarized light from the disk. This dilution decreases very significantly the polarization detected. For example $\beta$ Pictoris has been found to be 15\,per cent polarized when the star is hidden \citep{1991MNRAS.252P..50G} but through a whole aperture, the intrinsic polarization was measured to be only 0.2\,per cent \citep{2000ApJ...539..424K}. Hence, we might overlook debris disks if the mean polarization uncertainty in our survey is too high. In order to push forward and have better statistics, we perform in the next section a statistical analysis on a larger sample of objects.

\section{Statistical Comparison}\label{StatComp}


In order to obtain more robust results, we merged our observations with other ones. We used the \citet{1993A&AS..101..551L} catalogue which is an extensive polarization survey of precise measurements for 1000 stars closer than 50 pc. We note that his selection of stars was based on pre-$Hipparcos$ distances. In addition to his own measurements, \citet{1993A&AS..101..551L} compiled measurements from the literature that met his criteria. \cite{2013ApJ...768...25G}, based on $Spitzer$ MIPS and $Herschel$ PACS data, were able to reliably determine the presence of FIR excess for more than 550 nearby stars. By combining these two lists with our observations and keeping the measurement with the best S/N ratio for each object, we have the polarization and the occurrence of FIR excess for 223 stars. We include also HR~8799 although it is not in the DEBRIS sample since its distance is compatible, $< 46$ pc, the limit for A-type stars. The results for 18 stars with detected polarization ($P \geq 3\sigma_P$) are presented in Table~\ref{tab:3sigma}; those for the other stars are in Table~\ref{tablestars} in Appendix. 

Column~1 of Table~\ref{tab:3sigma} shows the star identification given by the DEBRIS and SONS surveys \citep{2010MNRAS.403.1089P,2012yCat..74031089P}, the first letter represents its spectral class and the number is a zero-padded running number increasing with distance in each subsample. These identifiers are referred to by the acronym UNS, standing for Unbiased Nearby Stars, as in the original SUNS survey names. Column~2 gives the name of the primary star; the choice of name is generally in the order of preference: HD, HIP, GJ, LHS, NLTT, TYC, PPM, CCDM, other catalogue name and 2MASS, following \cite{2012yCat..74031089P}. The other columns in Table~\ref{tab:3sigma} give the measured polarization and its uncertainty, the equatorial polarization angle of the polarization vector and its uncertainty. When the uncertainty of the measured polarization angle (given by eqn \ref{Stheta-eq}) is larger than the standard deviation of a completely random sample, $\pi / \sqrt[]{12} = 51.96\degr$, the polarization angle is undefined. Finally, the last five columns in Table~\ref{tab:3sigma} represent respectively the distance (pc) and its uncertainty, the ratio of the polarization to the distance ($P/d$), the presence of a disk according to FIR excess \citep{2013ApJ...768...25G}, the source of the data and the observation dates for OMM data. When we measured the polarization of a given star many times (multiple dates appear in the table), the result given is the weighted mean of these measurements. Table~\ref{tablestars} (Appendix \ref{Appen}) follows the same format than Table~\ref{tab:3sigma}  except that the two columns for the polarization angle contain "undefined values" for those stars with $P < 2\sigma_P$ for the OMM data or when the information was not available from \citet{1993A&AS..101..551L}. 

The column with the authors is from the \cite{1993A&AS..101..551L} paper with Simon (this paper) added where necessary. We also added \cite{2000A&A...362..978B} who observed about 10 stars from the DEBRIS list. Observations from OMM were made at a wavelength of 766 nm. The other stars compiled by \cite{1993A&AS..101..551L} were observed at various wavelengths between 400 nm and 600 nm depending on the observer. These differences in wavelength between observations was not relevant for \cite{1993A&AS..101..551L} because he was studying the distribution of interstellar dust in the solar neighbourhood, so, all observations in the visible were suitable for his purpose. These differences in wavelength may be significant for interpretation in our case since polarization may vary with wavelength across the visible. In our numerical simulations we computed the polarization for the $I$ band at 760 nm (see section \ref{Numresults}).

\subsection{Stars with 3$\sigma_P$ polarization detection \label{sub:3sigma}}

\begin{table*}
  \begin{center}
\caption{All stars observed at OMM and from the Leroy compilation with a polarization $P \ge 3 \sigma_P$.\label{tab:3sigma}}
\begin{threeparttable}
    \begin{tabular}{crrrrrrrccrl}
    \hline\hline
      UNS & HD & $P$ & $\sigma_P$ & $\theta(\degr)$ & 
$\sigma_\theta(\degr)$&$P/\sigma_P$&Distance&$P/d$&FIR\tnote{a}&Observers\tnote{b}&Date\tnote{c} \\
ID &&($10^{-5}$)&($10^{-5}$)&&&&(pc)&$(10^{-5}$pc$^{-1})$&excess&&\\
      \hline
K046&115404&152&43&8.4&7.4&3.5&11.095 $\pm$ 0.090&13.7 $\pm$ 3.9&?&SI&22 Jan\\
G013&101501&24&8&ind.\tnote{d}&ind.&3.0&9.602 $\pm$ 0.024&2.5 $\pm$ 0.8&N&BE&\\
G052&142373&15&5&ind.&ind.&3.0&15.894 $\pm$ 0.053&0.9 $\pm$ 0.3&N&PI WA LE SE&\\
F002&170153&38&12&ind.&ind.&3.2&8.032 $\pm$ 0.033&4.7 $\pm$ 1.5&N&TI WA&\\
F009&102870&26&8&ind.&ind.&3.3&10.928 $\pm$ 0.026&2.4 $\pm$ 0.7&N&TI SE BE MA&\\
F013&210027&10&3&ind.&ind.&3.3&11.719 $\pm$ 0.086&0.9 $\pm$ 0.3&N&PI WA SE BE&\\
F016&147584&26&7&ind.&ind.&3.7&12.113 $\pm$ 0.076&2.2 $\pm$ 0.6&N&TI MA&\\
F021&222368&12&4&ind.&ind.&3.0&13.716 $\pm$ 0.028&0.9 $\pm$ 0.3&N&SC PI MA BE&\\
F029&176051&24&6&ind.&ind.&4.0&14.881 $\pm$ 0.081&1.6 $\pm$ 0.4&N&WA\\
F032&7570&75&6&47.7&2.3&12.5&15.115 $\pm$ 0.055&5.0 $\pm$ 0.4&Y&KO TI SC&\\
F037&165908&15&5&159.9&9.6&3.0&15.660 $\pm$ 0.083&1.0 $\pm$ 0.3&Y&PI WA SE&\\
F040&76943&14&3&ind.&ind.&4.7&16.061 $\pm$ 0.172&0.9 $\pm$ 0.2&N&PI TI BE\\
F075&119756&90&30&ind.&ind.&3.0&19.407 $\pm$ 0.072&4.6 $\pm$ 1.5&N&MA&\\
A017&118098&29&8&ind.&ind.&3.6&22.724 $\pm$ 0.098&1.3 $\pm$ 0.4&N&TI BE MA&\\
A026&106591&15&5&ind.&ind.&3.0&24.688 $\pm$ 0.085&0.6 $\pm$ 0.2&N&PI TI BE&\\
A065\tnote{e}&137909&80&10&ind.&ind.&8.0&34.281$\pm$ 0.892&2.3 $\pm$ 0.3&N&LE&\\
A106&210049&20&4&ind.&ind.&5.0&41.592 $\pm$ 1.674&0.5 $\pm$ 0.1&N&SC&\\
A130&16555&22&6&ind.&ind.&3.7&45.538 $\pm$ 2.276&0.5 $\pm$ 0.1&N&SC&\\
      \hline \hline
    \end{tabular}  
  \begin{tablenotes}
  \item[a] Detection of FIR excess: yes (Y), no (N), no information or uncertain (?).
  \item[b] Abbreviations for the observers are, respectively, `SI' for this paper (Simon), `AP' for \cite{1968ApJ...151..907A}, `BE' for \cite{1959VeGoe...7..200B},`BM' for \cite{2000A&A...362..978B}, `HU` for \cite{1985A&A...152..357H, 1988ApJ...329..882H, 1990A&A...231..588H}, `KO` for \cite{1986A&AS...64..487K}, `KR` for \cite{1980A&AS...39..167K}, `LE` for \cite{1993A&AS..101..551L}, `MA' for \cite{1970MmRAS..74..139M}, `PI` for \cite{1977A&AS...30..213P},  `SC' for \cite{1976A&AS...23..125S}, `SE` for \cite{1970ApJ...160.1083S}, `TI' for \cite{1982A&A...105...53T} and `WA` for \cite{1968PASP...80..162W}.
  \item[c] Observations from OMM were obtained during the winter 2009-2010.
  \item[d] When the uncertainty on the polarization angle is larger than $\approx 52\degr$, its orientation is indefinite (see text). Also it is customary not to give the polarization angle when the polarization is considered to be too small to yield a reliable polarization angle. 
\item[e] 
As explained in section \ref{sub:3sigma}, there are two entries for this star, here and in Table \ref{tablestars}.  
\end{tablenotes}
\end{threeparttable}
\end{center}
\end{table*}

Table~\ref{tab:3sigma} shows that 18 objects have a polarization level at or above $3\sigma_P$. Two of them, HD~7570 and HD~165908, have a detected FIR excess. 

HD~165908, also known as 99 Herculis, is a binary system with F- and K-type stars at 16.5 AU from each other and with an imaged debris disk at 120 AU from the barycentre \citep{2012MNRAS.421.2264K}. Their preferred disk model is a ring of polar orbits that move in a plane perpendicular to the pericentre direction. The polarization angle is 158.9$\degr \pm 9.6\degr$, as computed directly from the \cite{1977A&AS...30..213P} data. This means that the observed polarization is almost aligned with the pericentre position angle, at 163$\degr$, which is more or less the expected orientation of the polarization vector, although it should vary slowly in time with the orbit of the binary with a period of 56.3 yr (see e.g., \cite{1978A&A....68..415B}, \cite{2005ASPC..343..389M}). \cite{1977A&AS...30..213P} did not mention the size of the aperture he used for his polarization observations. A diameter of 16\,arcsec includes most of the disk, at the distance of HD~165908. 

HD~7570 or $\nu$ Phenicis is a solar type star (F9 VFe) at 15 pc with confirmed IR excess \citep{2013ApJ...768...25G,2016ApJS..225...15C}, but no disk has been imaged yet in this system. \cite{2006ApJ...639.1166B} reported a $Spitzer$ IRS 30--34 $\mu$m excess, extending out to 70 $\mu$m. They estimated $L_{dust}/L_* = 4.3 \times 10^{-5}$ and modelled this excess with a single-ring disk between 11--12 AU with 10 $\mu$m dust grains at about 100 K. The disk mass is 1.3 $\times 10^{-6}$ M$_\oplus$ assuming silicate grains with a density of 3.3 g~cm$^{-3}$. Extrapolating the grain size distribution with a power of -3.5 out to 10-km diameter yields a mass of 0.042 M$_\oplus$. A submm detection of the system would help constrain the dust size and distribution and hence the total mass. This star has a strong polarization detection, with $P = 0.075$ per cent ($P/\sigma_P = 12.5$) at a polarization angle of $47.7\degr \pm 2.3\degr$ \citep{1986A&AS...64..487K}. Given the level of polarization and assuming single scattering in an optically thin disk, the disk should have a significant inclination and a projected major axis oriented near a position angle of $138\degr$.

There are 15 other stars with $3\sigma_P$ polarization detection but with no known IR excess. Such relatively low levels of polarization, $<P> = 0.023$\,per cent, can arise from various situations. Seven systems are known to be spectroscopic binary or multiple star systems which breaks the axisymmetry of the system, resulting in a non-null polarization. 
Also, it might be a property of some variable stars since five of them are photometric variables. Finally, it is also possible that their disks are too faint to be detected in the FIR. Here is an example of a photometric and polarimetric variable. HD 137909 ($\beta$ CrB) is a well-known $\alpha^2$ CVn magnetic variable star with a spectral type of A8 V SrCrEu \citep{2010MNRAS.403.1089P,2003AJ....126.2048G}. The broad-brand continuum linear polarization traces variations of the magnetic field component perpendicular to the line of sight as the star rotates \citep{1993A&AS..101..551L} with an amplitude of order $\sim 10^{-3}$. \cite{2010MNRAS.403.1089P} had two entries for CCDM 15278+2906A and CCDM 15278+2906B as A065 in their Table 7, and also their Table in the Vizier catalogue. Both of them correspond to HD 137909 ($= \beta$ CrB). Its polarization was detected by Leroy but not by Simon. This led us to two different entries for this star, the detection by Leroy in Table~\ref{tab:3sigma} and the non detection by Simon in Table~\ref{tablestars}. This has no significant effect on our statistics below.

Last, HD~115404 is a $3\sigma_P$ OMM detection, but there is no information about a possible FIR excess. It is a binary system which might explain the polarization detection. It consists of a K1\,V star and a M1 dwarf and it has no known disk. As mentioned above, it is one detection in a sample of slightly more than one hundred stars (the OMM sample discussed in section \ref{OMM-results} above), just the single detection expected statistically for such a sample. The polarization of this star should be confirmed by new observations. 

\subsection{Analysis \label{analysis}}

\begin{figure}
\centering
\includegraphics[angle=0,origin=bl,width=0.95\columnwidth]{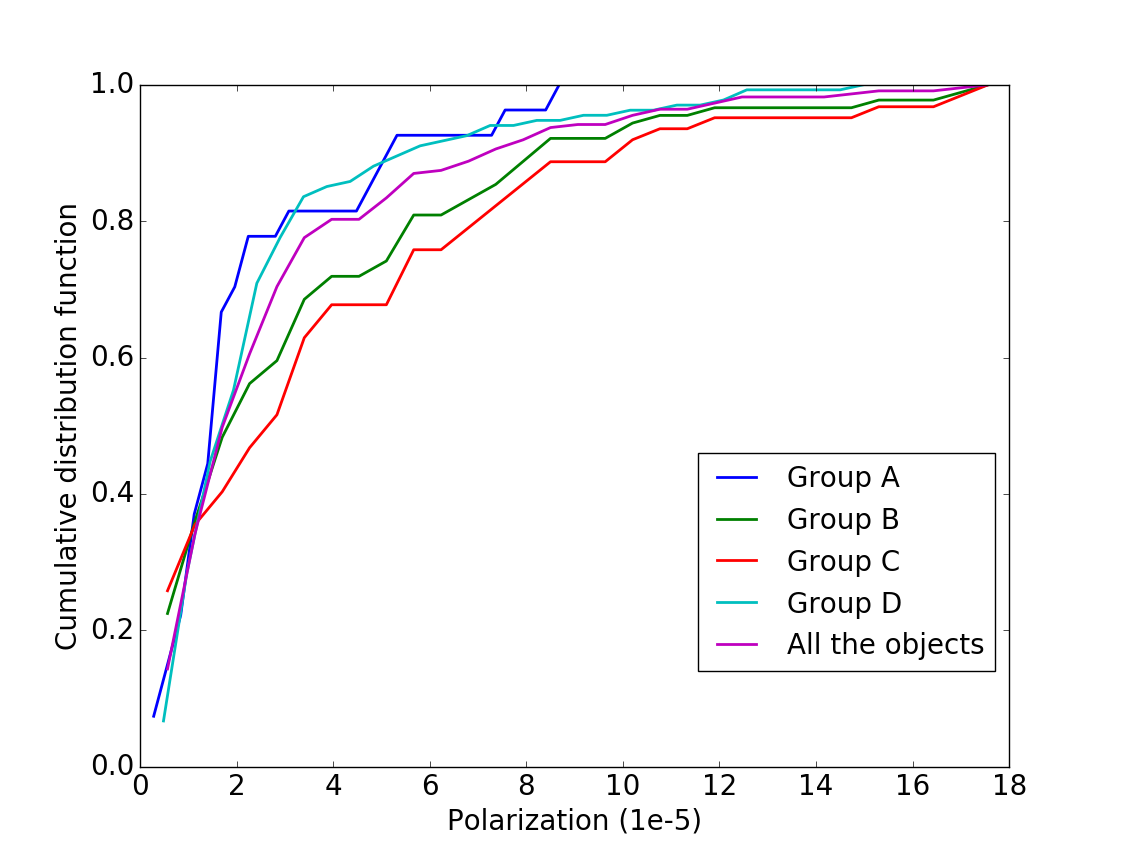}
\caption{Comparison of the polarization cumulative distribution function (CDF) for each group of stars defined by their presence of FIR excess. The fractional number of stars with observed polarization smaller than a given value is given as a function of the polarization. Groups A, B, C and D as defined in section \ref{analysis} are  represented by the colours dark blue, green, red and light blue respectively. The purple line corresponds to the distribution for the entire sample.}
\label{figCDF}
\end{figure}

We divide our sample of 223 stars into 4 groups to compare their statistical properties. The A group contains objects with confirmed FIR excess as shown by Y in Tables \ref{tab:3sigma} and \ref{tablestars}. The group B contains group A stars plus those suspected to have a FIR excess or for which no confirmation is available yet (Y + ? in Tables \ref{tab:3sigma} and \ref{tablestars}). Group C contains only those suspected to have a FIR excess or for which no confirmation is available yet (? in Tables \ref{tab:3sigma} and \ref{tablestars}). And the last group, D, contains stars known for not having a FIR excess (N in Tables \ref{tab:3sigma} and \ref{tablestars}). The cumulative distribution function for the polarization of each group and of the total sample is given in Figure~\ref{figCDF}.

\begin{table*} 
\begin{center}
\caption{Summary of the polarization statistics for each group of stars. The description of the samples and the meaning of each column are given in the text. The units for the polarization and uncertainties are $10^{-5}$.}
\label{tab:meanpola}
\begin{threeparttable}
\begin{tabular}{l|rcccccc}
Sample & $N$ & $N_{P>3\sigma} $ &det. rate\tnote{a} & $<P>$\tnote{b} & $<\sigma_P $>\tnote{c} & 
$\zeta_P $\tnote{d}& $\zeta_{\sigma_P} $\tnote{e}\\ 
\hline
\hline
A (Y)   &  28 &  2 & 7$^{+8}_{-2}$ \% & $23$ & $23$ & $14$ & $11$\\
B (Y+?) &  88 &  3 & $3 \pm 2 \%$ & $33$ & $38$ & $28$ & $19$\\
C (?)   &  60 &  1 & $2 \pm 2 \%$ & $38$ & $42$ & $34$ & $18$\\
D (N)   & 135 & 15 & $11 \pm 3 \%$& $25$ & $26$ & $17$ & $16$\\
All 	& 223 & 18 & $8 \pm 2 \%$& $28$ & $32$ & $21$ & $18$\\
\hline
\hline
\end{tabular}
\begin{tablenotes}
\item[a] Given the small numbers involved, the binomial distribution is appropriate. We computed 1$\sigma$ uncertainties following \cite{2003ApJ...586..512B}. 
\item[b] Mean polarization of the groups.
\item[c] Mean uncertainties of the groups.
\item[d] Square root of the variance of the polarization of the groups.
\item[e] Square root of the variance of the uncertainties of the groups.
\end{tablenotes}
\end{threeparttable}
\end{center}
\end{table*}

For each group we computed different parameters presented in Table~\ref{tab:meanpola}. The columns in the table show: group identification and in parentheses the FIR excess status, the number of stars in each group, the number of stars with a detected polarization, i.e. with $P \geq 3\sigma_P$, the detection rate and its uncertainty based on the two previous columns, the (unweighed) average polarization $<P>$, the (unweighed) average uncertainty of the measurements $<\sigma_P>$, the standard deviation of the polarization measurements $\zeta_P $, and the standard deviation of the uncertainties $\zeta_{\sigma_P}$. The first point we can observe is even if the average polarization differs between the different groups, in particular A, C and D, these differences are not significant because they are smaller than $<\sigma_P>$ and $\zeta_P $. Checking the mean of the measurement uncertainty, $<\sigma_P>$, and the statistical uncertainty on these, $\zeta_{\sigma_P}$, we can see the same variation pattern. Finally no significant differences have been found between the different groups except for the measurement precision. We note that the sample about which we have less information (C) is also the less precise one. 

\begin{table}
\centering
\caption{Results of the Kolmogorov--Smirnov (KS) tests between different debris-disk groups with respect to polarization detection.}
\begin{tabular}{l|ccc}\label{KStest}
 & A vs. C & A vs. D & C vs. D\\ 
\hline
\hline
KS coefficient &  0.310  & 0.159 & 0.260 \\ 
p-value &  0.042  & 0.581 & 0.005 \\
\hline
\hline
\end{tabular}
\end{table}

To pursue the analysis further, we performed a Kolmogorov--Smirnov (KS) test for different combinations of the stellar groups. We tested the "null hypothesis", $H_0$: "The two groups of stars tested through polarization data come from the same population of stars," i.e. the two groups of stars tested have the same polarization Cumulative Distribution Function (CDF). The KS statistics, $D_{KS}$, obtained with the KS test represents the maximum distance between the CDF of two groups being compared. We set a confidence level $\alpha$ and for each statistical test compute a critical value $C_{\alpha}$\footnote{Tables can easily be found in literature.} depending on $\alpha$ and the number of elements in the list. Then, we can reject $H_0$ with a confidence level of $\alpha$ if $D_{KS}$ is greater than $C_{\alpha}$. In Table \ref{KStest}, KS results are generally low. We have to choose a low level of confidence ($\alpha=0.1$), to reject $H_0$ for A versus C (where $C_{\alpha}\simeq 0.28$) and C versus D (where $C_{\alpha}\simeq 0.19$). So, we can reject $H_0$ for A versus C and C versus D but not for A versus D. Since we used a low level of confidence, checking $H_0$ with another parameter would help to confirm our conclusions. The KS test also computes the p-value which represents the believability that $H_0$ can be rejected\footnote{If $p\leq \alpha $, $H_0$ may be rejected; but if $p> \alpha$ we cannot conclude about the validity of $H_0$}. With those results, we notice that the null hypothesis can be rejected for A vs. C and for C vs. D. In conclusion, the probability that the samples have consistent CDF's statistics is very low and we can confidently reject $H_0$ for A versus C and C versus D tests. For the A versus D test, we cannot reject the null hypothesis, i.e., they may come from the same population of stars. 

To summarize, we can conclude that sample C is different from samples A and D but that sample A may be consistent with D. So, the conclusion of the KS test is unexpected: the two samples we would legitimately assume as different could possibly match; and the samples that should match are, in fact, different. This is consistent with Figure \ref{figCDF} where group C presents the largest differences with the other groups. 

\cite{2000A&A...362..978B} made a similar comparison for 61 stars and they found a larger polarization for Vega-like stars (i.e., stars with debris disks) than for normal field stars. They observed 27 stars with known IR excess and selected measurements for 34 additional stars with IR excess from the \cite{2000AJ....119..923H} catalogue. Uncertainties on measurements in the Heiles catalogue are $\approx 0.1$ per cent. The distances for these stars are significantly greater than those used here; only about 10 of them are in the DEBRIS list.

\subsection{Conclusion}

When we compare the groups with and without FIR excess (A vs. D), we find that there is no statistically significant difference for their $3\sigma$-polarization detection rate, nor for the shape of their cumulative polarization distribution functions. This is also confirmed by the KS test. The group that differs the most from the other ones is group C, the objects without FIR excess confirmation.

These results seem to indicate that there is no correlation between the presence of a disk and the observed level of polarization. But, we should not forget that the great majority of our polarizations are below the $3\sigma$ level and that the detection rate obtained is smaller than the fraction of stars with debris disks, around 20 per cent. This last statement, a polarization rate smaller than the fraction of stars with debris disks, would  be expected since polarization also depends on the orientation of the disks with respect to the observer. However, it is also likely that some debris disks are not detected because their IR emission is too low but their polarization can nevertheless be detected. If this is the case, it should depend on the sensitivity of the IR detectors relative to that of the polarimeters used. This goes against our equation \ref{Pr-rel} (see section \ref{Cold-disks}), but is not impossible.  So, we cannot conclude firmly about the link between disks and polarization based on observations only. 

In order to complete our investigation about this link we now turn to analytical models and simulations.

\section{Models}\label{Models}
	

A conclusion of the previous sections is that few systems have a linear polarization above the typical threshold of $P\sim 10^{-4}$ integrated through an aperture. In particular, we see in Table~\ref{tab:3sigma} that the polarization of detected stars ranges from about 1 to a few $\times 10^{-4}$. 
In the following, we will consider $P = 10^{-4}$ as a polarization limit and investigate the constraints it imposes and its implications for the mass of circumstellar dust in these systems. This polarization threshold can be considered as a practical limit between classical and high-precision polarimeters and will be used as a reference here. 

\subsection{Analytic approach}\label{AA}

\begin{figure}
\centering
\includegraphics[angle=0,origin=bl,width=0.95\columnwidth]{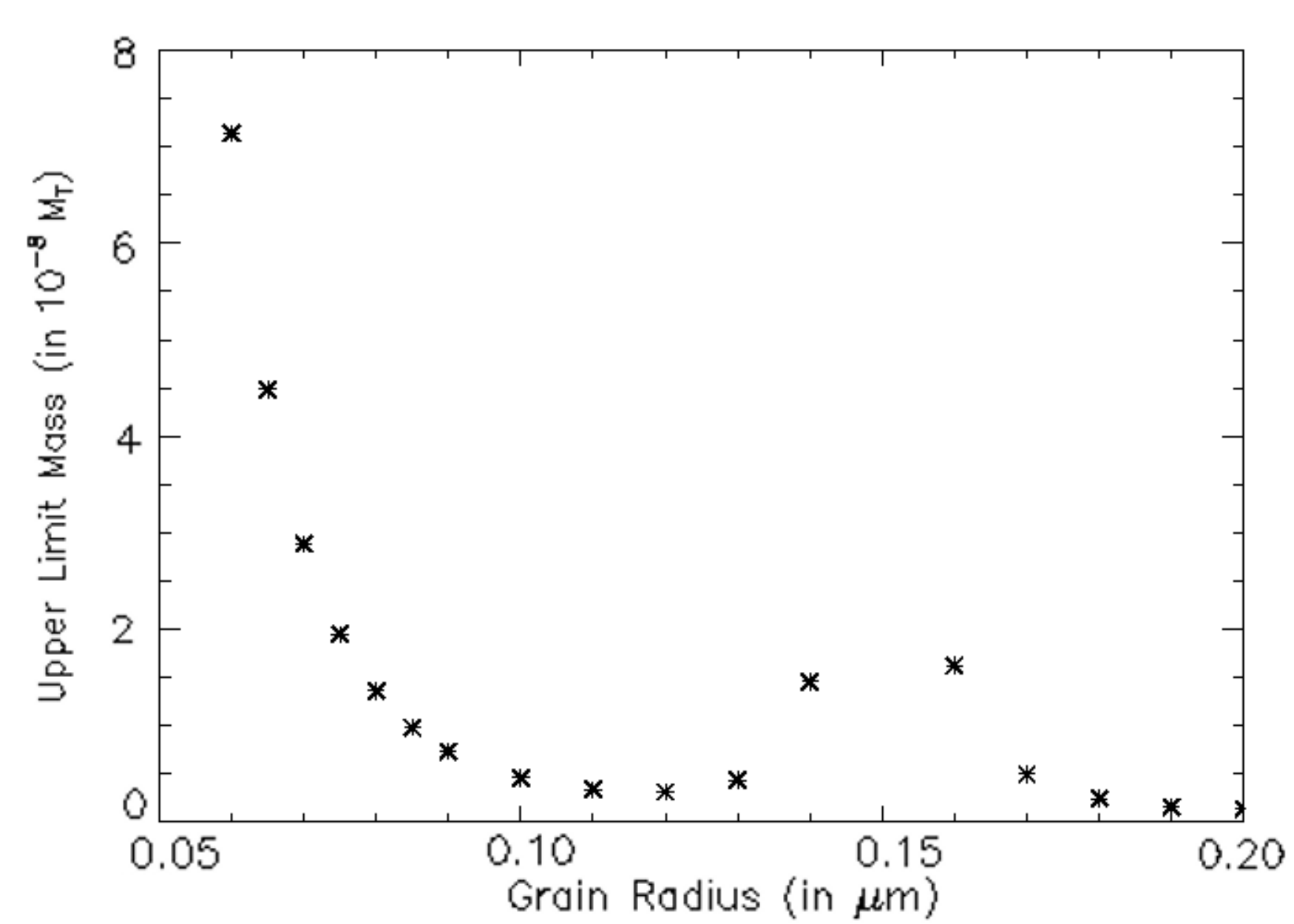}
\caption{Mass of the disk determined as a function of the grain radius, for an assumed observed polarization of 0.1 per cent. The mass is given in terrestrial mass, (M$_{\rm T}$ = M$_{\oplus}$). Other assumptions are given in the text. There is no symbol for a radius of 0.15 $\mu$m due to the very low polarization induced by grains of this size. This leads to a very high disk mass, larger than the scale of this figure.}
\label{fig4}
\end{figure}

We consider a toy model in which a cloud of identical dust grains is placed at a unique distance and unique azimuth from the star. Such a configuration mimics the case of an isolated blob of dust, but we acknowledge that this most likely does not correspond to any realistic situation but a scenario which maximises the polarization for a given dust mass. The scattering angle of $90\degr$ maximizes approximately the polarization signal, and yields an estimate of the minimum dust mass necessary to reach a $P=10^{-4}$ polarization level. Assuming unpolarized incident light, the degree of linear polarization $P$, measured by the observer with aperture polarimetry, is given by
\begin{equation}
P=\frac{P_{\rm dust}I_{\rm c}}{I_{\star}+I_{\rm c}+I_{\rm th}},
  \label{P-eq}
\end{equation}
where $P_{\rm dust}$ is the degree of linear polarization for scattering by one grain, $I_{\rm c}$ the intensity of stellar light scattered by the dust cloud, $I_{\star}$ the stellar intensity and $I_{\rm th}$ the thermal grain emission at the considered wavelength. The degree of linear polarization of one grain ($P_{\rm dust}$) and the scattered intensity of the cloud ($I_{\rm c}$) are given by (see e.g., \cite{1998asls.book.....B}):
\begin{eqnarray}
P_{\rm dust}=\frac{i_1 - i_2}{i_1 + i_2} , 
  \,\,\,\, I_{\rm c}=N I_{\star}\frac{i_1+i_2}{2k^2r^2}=N I_{\star}y, \label{Ic-eq}
\end{eqnarray}
where $i_1$ and $i_2$ denote the scattered-light fractional intensities polarized perpendicular and parallel to the scattering plane, respectively, $r$ is the distance to the star, $k$ the wavenumber and $N$ the number of grains. This number of grains is linked to the dust mass $M$ of the disk by $M = \rho V N$ with $\rho$ the grain density (we used the value of 3~g~cm$^{-3}$ for astronomical silicates) and $V$ the volume of one grain.

Combining equations~\ref{P-eq} and \ref{Ic-eq}, and anticipating that $I_{\rm c}+I_{\rm th} \ll I_{\star}$, we obtain the dust cloud mass
\begin{eqnarray}
M \simeq \frac{P \rho V 2k^2r^2}{i_1-i_2}.
\label{Mdust-eq}
\end{eqnarray}
 
Combining equations~(\ref{P-eq}) and (\ref{Ic-eq}), we then have
\begin{equation}
P=P_{dust}N \left[\frac{y}{1+Ny+I_{\rm th}/I_{\star}}\right].
\label{Ptot-eq}
\end{equation}
 
The polarization induced by debris disks will decrease approximately as the square of their distance from the star, as can be seen in equation (\ref{Ptot-eq}) and remembering that 
$y \propto r^{-2}$ (see equation \ref{Ic-eq}). Note that the 
last two terms in the denominator are negligible since $I_{\rm c}+I_{\rm th} \ll I_{\star}$ in the visible and NIR (see e.g., \cite{2014AJ....148...59S}). Therefore, for $a~given~mass$ (or number of dust grains, $N$), this survey is more sensitive to hot debris disks such as the zodiacal cloud in our solar system. In this simple model, we assume arbitrarily that all grains are at 1~au. This is a representative distance such that grains are warm enough and such that polarimetric measurements are sensitive. 

\cite{2002ApJ...578.1009F} estimated that grain radii in the zodiacal cloud range from 0.01~$\mu$m to 1~cm, decreasing exponentially. For our simple model, we use a grain radius of 0.1~$\mu$m. We modelled the grains with astronomical silicates and took the complex refractive index from \cite{1985ApJS...57..587D} at 0.870 $\mu$m: $m~=~n+ik$ with $n=1.71$ and $k=0.0297$. Using Mie scattering theory, we computed numerically the Van de Hulst intensities, $i_1$ and $i_2$, for a given grain radius and scattering angle. Figure \ref{fig4} shows how other grain sizes will change the total mass. 

With all these assumptions, we find a disk mass of 0.1~$\mu$m-sized grains of 4.5 $\times$ 10$^{-10}$ M$_{\oplus}$ for $P = 10^{-4}$. Note that Figure \ref{fig4} is scaled for $P = 10^{-3}$. As mentioned above, this model finds the minimum mass of grains with a given radius that can produce a polarization $P = 10^{-4}$. This is the case because we use a single position at 1 au for all the scattering dust grains and also a single scattering angle of $90\degr$ which is near the maximum of the scattering phase function. If conditions are less favourable, a larger mass will be necessary to reach the same polarization. Since the zodiacal cloud mass has been evaluated at between 0.33 -- 1.8 $\times$ 10$^{-9}$ M$_\oplus$ by \cite{2002ApJ...578.1009F}, and at $\sim$ 8 $\times$\,10$^{-9}$ M$_\oplus$ by \cite{2010ApJ...713..816N} based on dynamical models and IRAS data, we indeed are at the limit of sensitivity of the instrument, if extrasolar hot debris disks resemble the zodiacal cloud.

Single scattering by Mie particles in optically thin envelopes around a star has been considered in the past and can be applied in the context of our paper. \cite{1987ApJ...317..231B} studied analytically the properties of different geometries. For an azimuthally symmetric but otherwise arbitrary density distribution, the polarization is proportional to $\sin^2 i$ as long as grains are relatively small compared to the wavelength, i.e., $x = ka = 2\pi a/\lambda < 2.0$, where $a$ is the grain radius and $i$ is the inclination of the disk. For a plane disk, the polarization is given by \citep{1987ApJ...317..231B}:
\begin{equation}
  P \propto \frac{15}{4}[N'\sin^2 i]\overline{F}_{22}, 
  \label{Pdisk-eq}
\end{equation}
where $N'$ is an integral over the radial density distribution in the disk, and $\overline{F}_{22}$ is real and depends on the scattering phase function \citep{1982MNRAS.200...91S}. $\overline{F}_{22}$ also determines the wavelength dependence of the polarization. The polarization angle is usually perpendicular to the disk, except for a possible change by $\pi/2$ when $\overline{F}_{22}$ changes sign. In all cases, we see that there is a strong, sine square, dependence on the inclination. This dependence on inclination is responsible for the non detection of polarization for a large fraction of stars which have been detected by IR surveys (see section \ref{analysis}), simply because they do not show a favourable configuration towards Earth to generate a significant polarization.

\subsection{Simulations of dust belts}

Complementary to our analytical toy model, we performed extensive simulations with MCFOST \citep{2006A&A...459..797P}, a Monte Carlo Radiative Transfer (MCRT) code. These results enable us to investigate the influence of many parameters on the polarization of unresolved objects. As earlier, we assume a detection threshold of $10^{-4}$ for linear polarization.

We then computed 2-D maps for all Stokes parameters at $\lambda=0.76\,\mu$m (which is approximately the central wavelength of the bandpass used by Beauty and the Beast), with the size of the map corresponding to the extent of the disk itself. 

The most important parameter to investigate is the distance between the inner edge of the disk and its star. As explained earlier, closer the disk is located, higher the polarization will be. But the large field of view of aperture polarimeters cannot be used to constrain belt locations. As a consequence, the spectral type of the star is the parameter that constrains the location of the disk: smaller and colder the star is, closer the disk can be. As a consequence, we tested 4 spectral types as shown in Table~\ref{tab:stars}.

\begin{table}
\caption{Stellar data used in the calculations. The assumed synthetic spectral distribution of each star comes from NextGen simulations \citep{1999ApJ...512..377H}. Data in the table are from \citet{1973asqu.book.....A}.}
\label{tab:stars}
\centering
\begin{tabular}{c|ccccc}
Spectral Type & $R/R_\odot$ & $M/M_\odot$ & $T$ (K) & $\log g$ & $L/L_\odot$ \\ 
\hline
\hline
A0 & $2.40$ & $2.90$ & $9790$ & $4.13$ & $47$ \\ 
F5 & $1.30$ & $1.40$ & $6650$ & $4.36$ & $3$ \\ 
K0 & $0.85$ & $0.79$ & $5150$ & $4.48$ & $0.46$ \\
M2 & $0.50$ & $0.40$ & $3520$ & $4.64$ & $0.032$ \\ 
\hline
\hline
\end{tabular}
\end{table}

In order to compare disks, we use the temperature as a key parameter to set the inner and outer radii of each disk, mimicking roughly those from the solar system\footnote{Exozodiacal, asteroidal and Kuiper belts are also called hot, warm and cold belts.}.

For hot disks the inner edge, $r_{\rm in_{\rm hot}}$, is set at the dust sublimation radius crudely estimated assuming blackbody emission/absorption dust properties and thermal equilibrium:
\begin{eqnarray}
\label{eq:rsub}
R_{\rm sub}=\sqrt{\frac{L_{\star}}{\pi\sigma_s}}\frac{1}{4T_{\rm sub}^2}, 
\end{eqnarray}
where $L_{\star}$ is the stellar luminosity, $\sigma_s$, the Stefan-Boltzmann constant, and $T_{\rm sub}$ the sublimation temperature assumed equal to $1500$\,K, representative of typical silicate grains. Other inner radii are set to match typical temperatures: 300 K for $r_{\rm out_{hot}}$ and $r_{\rm in_{warm}}$, 50 K for $r_{\rm out_{warm}}$ and $r_{\rm in_{cold}}$ and finally 40 K for $r_{\rm out_{cold}}$. The radii computed for each representative star are shown in Table~\ref{tab:geometry}. A typical mass for each one of these disks is given in Table~\ref{tabpeak}. These masses come from the literature \citep{2006A&A...452..237A,2006ApJ...639.1166B,2005ApJ...635..625N,2007A&A...475..243D,2007ApJ...660.1556R,2003MNRAS.342..876W,2002MNRAS.334..589W,2013A&A...555A.146L} and were obtained from observations and/or disk modelling. We also used more recent information, presented in the second part of Table~\ref{tabpeak} for various spectral types from B8 to M2, but only 2 M-type stars are included (cold belts). Values for hot belts come from a reanalysis with detailed models by \cite{2017MNRAS.467.1614K} of published interferometric data. Warm belt values come from a compilation of $Spitzer$ (63) and $WISE$ (11) detected disk measurements by \cite{2016ApJ...826..171G}. We note that the average of the $WISE$ masses is about 50 times smaller than the average of only the $Spitzer$ data. Finally, data for cold belts come from the SONS survey \citep{2017MNRAS.470.3606H}. 

\begin{table}
\caption{Values of inner ($r_{\rm in}$) and outer ($r_{\rm out}$) radii for the belts around four representative stars computed from temperatures given in the text.}
\centering
\begin{tabular}{ccccc}
\multicolumn{2}{c}{ } & Hot & Warm & Cold \\ 
\hline 
\hline 
A0 & $r_{\rm in}$ [AU] & $0.236$ & $5.9$ & $212$ \\ 
 & $r_{\rm out}$ [AU] & $5.9$ & $212$ & $332$ \\ 
\hline  
F5 & $r_{\rm in}$ [AU] & $0.059$ & $1.49$ & $53.7$ \\ 
  & $r_{\rm out}$ [AU] & $1.49$   & $53.7$ & $83.8$ \\ 
\hline  
K0 & $r_{\rm in}$ [AU] & $0.023$ & $0.584$ & $21.0$ \\
  & $r_{\rm out}$ [AU] & $0.584$  & $21.0$ & $32.8$ \\ 
\hline  
M2 & $r_{\rm in}$ [AU] & $0.020$ & $0.502$ & $18.1$ \\
  & $r_{\rm out}$ [AU] & $0.502$  & $18.1$ & $28.2$ \\
 \hline
 \hline
\end{tabular}
\label{tab:geometry}
\end{table}

We studied the influence of the power-law grain size distributions. We assumed spherical, amorphous silicate grains [see \cite{1985ApJS...57..587D} for the optical constants], with sizes going from $a_{\rm min}$\,\,$\epsilon$  \{$10^{-1}, 1, 10$\} $\mu$m to $a_{\rm max}$ fixed to $1$ mm. Tiny grains dominate the scattering process because of their higher relative cross sections. Selecting a different $a_{\rm min}$ is equivalent to adding or removing tiny dust grains: it leads to an increase or a decrease of the scattering efficiency for a given total dust mass. 

The same effect is observed as we varied the index of the exponent in the power law of the grain size distribution: $\kappa = 3, 3.5$ and $4$. Finally, the effects of some geometrical factors were studied such as the surface density profile also defined as a power law, with index $\alpha=-3.5, -2$ and $-0.5$, and the thickness of the disk, an important geometrical factor, given by a gaussian vertical profile with a scale height defined as $H/r = 0.05$.

\begin{table} 
\begin{center}
\caption{Typical mass in M$_\oplus$ unit for each disk.}
\label{tabpeak} 
\begin{threeparttable}
\begin{tabular}{c|ccc}
 & Hot belts & Warm belts & Cold belts \\
 \hline 
 \hline
A0 & $2\times 10^{-6}$ & $3\times 10^{-4}$ & $1\times 10^{-2}$ \\ 
F5 & $3\times 10^{-7}$ & $2\times 10^{-4}$ & $4\times 10^{-3}$ \\ 
K0 & $3\times 10^{-8}$ & $6\times 10^{-5}$ & $3\times 10^{-3}$ \\ 
M2 & $4\times 10^{-9}$ & $2\times 10^{-5}$ & $1\times 10^{-3}$ \\ 
\hline
$<M>\tnote{a}$ & $2\times 10^{-9}$ & $8\times 10^{-5}$ & $9\times 10^{-2}$ \\
Range & $(0.2 - 4) \times 10^{-9}$ & $1\times 10^{-7} - 1\times 10^{-3}$ & $2\times 10^{-4} - 0.4$ \\
$N$ & 9 & 74 (63 + 11) & $46$ \\
\hline
\hline
\end{tabular}
\begin{tablenotes}
\item[a] This section of the table presents the average mass, the range of masses covered by the sample and the number of stars in the sample, $N$. The sample includes stars of spectral types A, F, G and K, plus a few B8 and M-types. See text for additional information. 
\end{tablenotes}
\end{threeparttable}
\end{center}
\end{table}

Finally we studied the influence of disk asymmetries. Debris disks are present in evolved planetary systems, and $Kepler$'s results have shown that there is a significant fraction of stars with planets. If planets are sufficiently massive, they will affect significantly the disk geometry as shown by \citet{2014A&A...563A..72F} for example. One of the most important effects is to break axial symmetry, creating elliptical orbits or ellipses instead of circular rings. With this geometry, one section of the ellipse will be closer to the star than the other ones. When the belt is a circular ring, every location along the ring produces the same linear polarization fraction as seen  pole-on, and the polarization from each location is cancelled by another location at $90\degr$ from it. The result is a net zero polarization. But this is no longer the case when axial symmetry is broken as polarization vectors will not cancel out. Our purpose here is to obtain a rough quantitative estimate of this effect on polarization and therefore on our mass detection threshold. The most important effect of this symmetry breaking is the differential amount of energy coming from the star on the disk. To mimic this effect we just move the star from the centre of the ring that we still model as a circle. Eccentricity is typically around $0.3$ to $0.4$ such as for example Fomalhaut \citep{2012ApJ...750L..21B}. To cover this range of eccentricities we move the star from the centre of the disk by respectively 0 per cent, 30 per cent and 60 per cent of the inner radius of the belt.

Combining all those elements, 4 spectral types, 3 disk types, 3 minimum grain sizes, 3 power-law exponents for the grain size distributions and 3 exponents for the surface density distributions and 3 stellar offsets, yields a computational grid of 972 disk models. The FITS file produced by each model has 12 relevant images for 3 disk inclinations and 4 disk azimuths (described below in \ref{Numresults}),  leading to 11\,664 elements to analyse. In summary, for each combination of spectral type and type of disk, there are 4 parameters considered (see Table~\ref{tabparam}). And each one of these models includes different inclinations and azimuths. 

\begin{table} 
\begin{center}
\caption{Values of the parameters used in the computational grid. }
\label{tabparam}
\begin{threeparttable}
\begin{tabular}{l|ccc}
Parameter name\tnote{a} & minimum value & mid value & maximum value \\
 \hline 
 \hline
$a_{\rm min}$ ($\mu$m) & $0.1$ & $1$ & $10$ \\
$\kappa$ & $3$ & $3.5$ & $4$ \\
$\alpha$ & $-3.5$ & $-2$ & $-0.5$ \\
Star's position & $0$ & $0.3 r_{\rm in}$ & $0.6 r_{\rm in}$\\
\hline
\hline
\end{tabular}
\begin{tablenotes}
\item[a] The first two parameters specify the grain size distribution: minimum grain size and power-law index. $\alpha$ is the surface density exponent of the disk and the stellar position is used for simulating eccentric disks (see Fig. \ref{belts2}).
\end{tablenotes}
\end{threeparttable}
\end{center}
\end{table}

For each model we compute the dust mass consistent with our detection threshold, $P = 10^{-4}$. For the computations, we adopt a fiducial distance of $10\,$pc from the observer. We note that the degree of polarization does not depend on this assumption: the polarized and total intensities and their ratio all scale as the square of the distance. We do not consider interstellar polarization in our calculations, but will discuss its effects on observed debris disks later in section \ref{Discussion}.

\subsection{Results}
\label{Numresults}

\begin{figure}
\centering
\includegraphics[width=\columnwidth,trim = .5cm 0.5cm 2cm 0.5cm, clip]{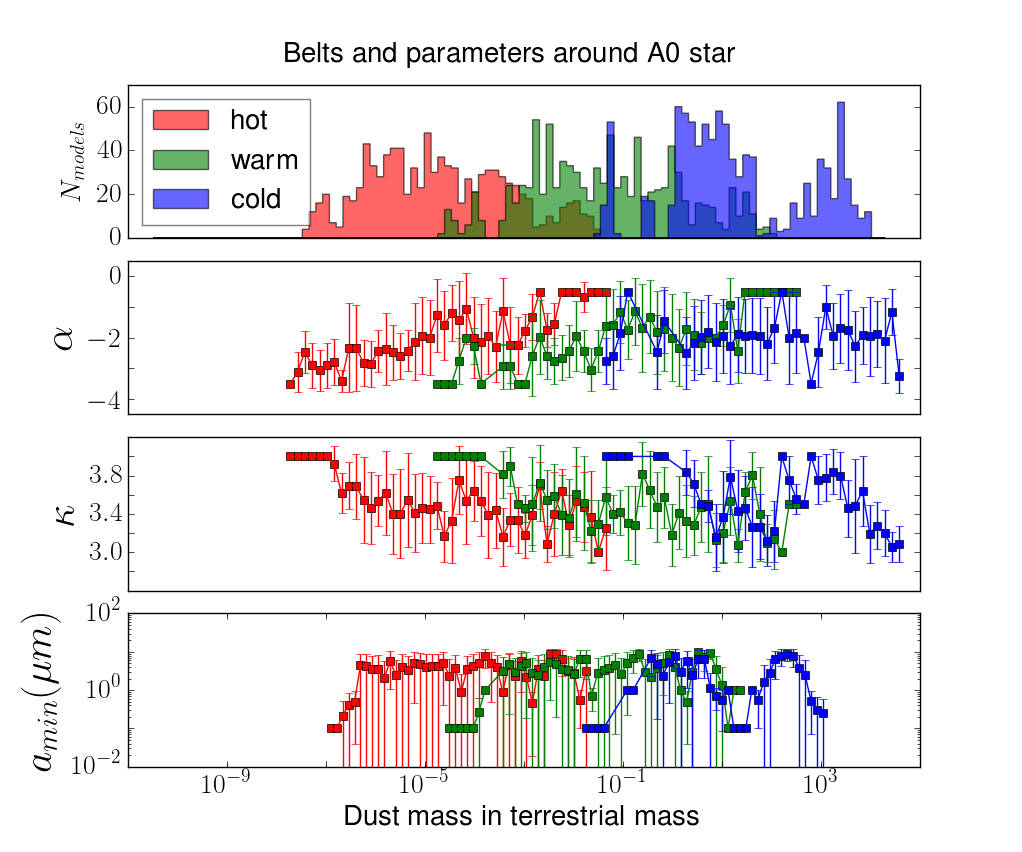}
\end{figure}

\begin{figure}
\centering
\includegraphics[width=\columnwidth,trim = .5cm 0.5cm 2cm 0.5cm, clip]{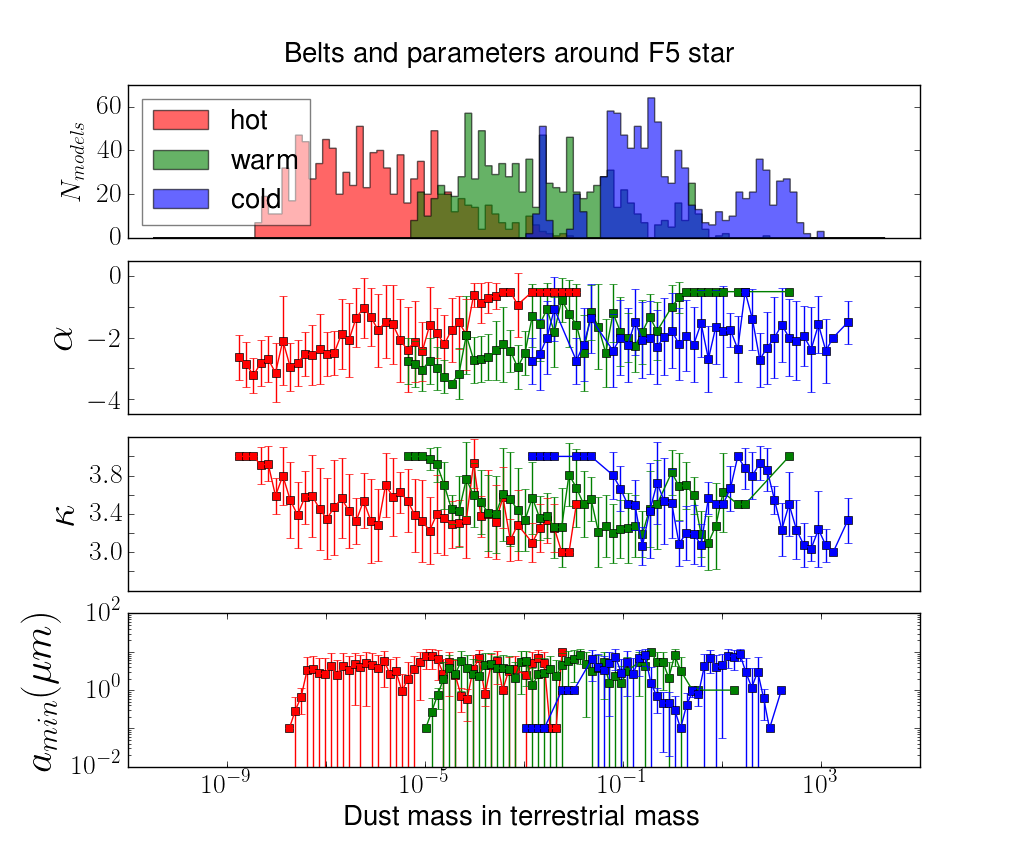}
\end{figure}

\begin{figure}
\centering
\includegraphics[width=\columnwidth,trim = .5cm 0.5cm 2cm 0.5cm, clip]{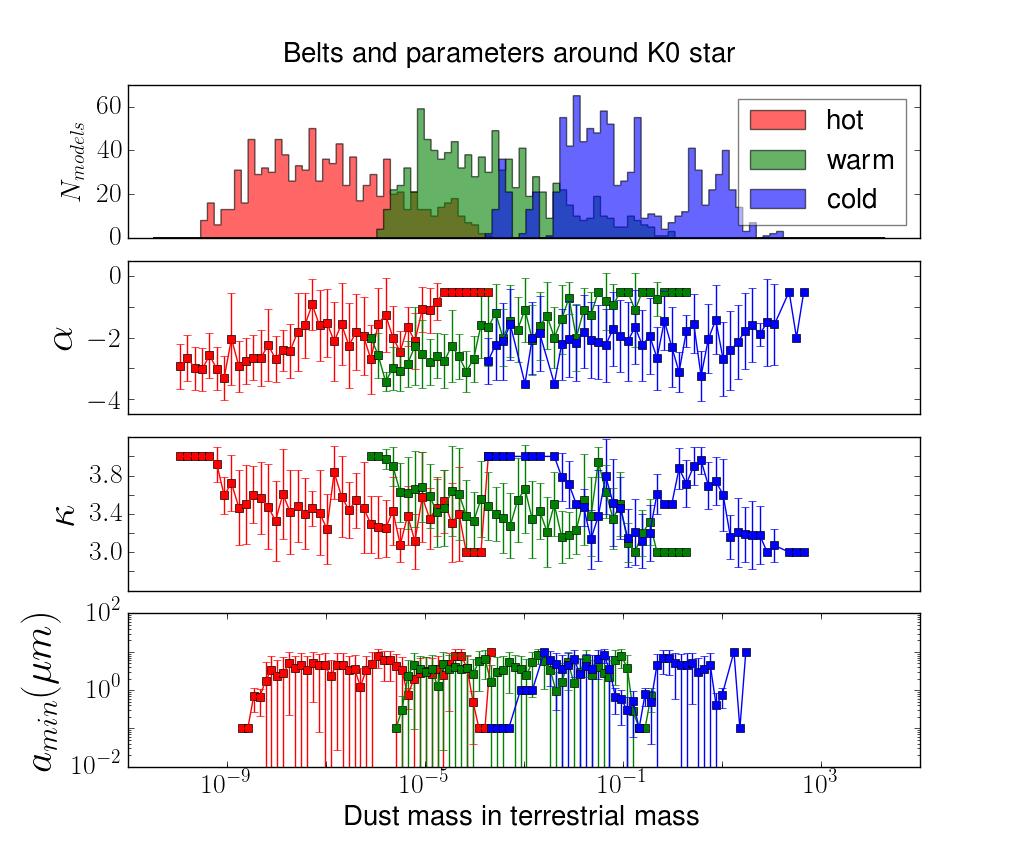}
\end{figure}

\begin{figure}
\centering
\includegraphics[width=\columnwidth,trim = .5cm 0.5cm 2cm 0.5cm, clip]{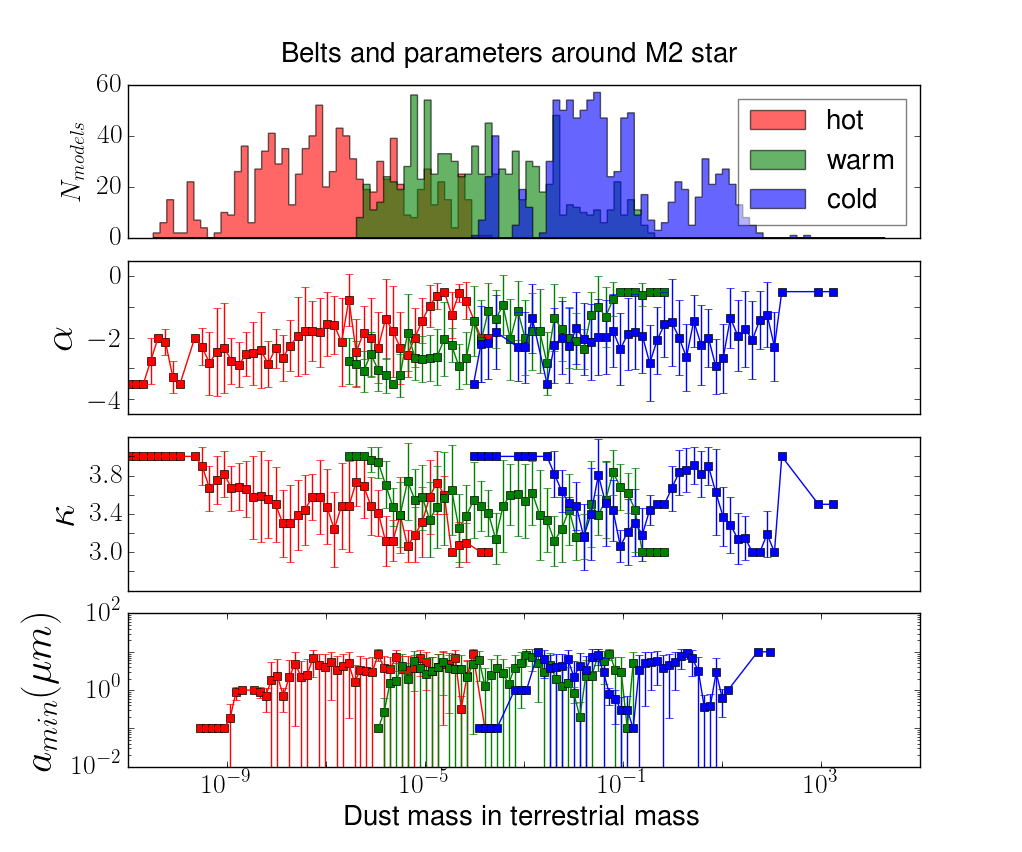}
  \caption{Histograms representing the number of simulations sorted by the mass of the minimum detectable mass of their disks. All 972 disk models are used, including the three stellar positions for simulating eccentricity. Underneath we plot the mean values in each mass bin for the surface density exponent $\alpha$, the grain-size  distribution power-law index $\kappa$, and the minimum size of the grains, $a_{min}$, for the 4 spectral types and 3 belts considered (see Table~\ref{tabparam}). The error bars correspond to 1 $\sigma$ and vary according to the number of simulations in each mass bin. }
  \label{belts}
\end{figure}

For each star and set of parameters MCFOST computes a synthetic cube of images of the Stokes parameters $I$, $Q$, $U$ and $V$ at $\lambda = 0.76\,\mu$m. These image cubes are computed for four different azimuthal angles ($\phi = 0\degr, 30\degr, 60\degr$ and $90\degr$) and three inclination angles ($i =0\degr$, $30\degr$ and $90\degr$). Finally, a simple python algorithm computes the total integrated linear polarization of the synthetic pictures.

We assume optically thin disks, i. e. the total polarization is proportional to the disk mass, according to equation (\ref{Ptot-eq}) above. Therefore we can scale the mass to make it correspond to the degree of linear polarization $P=\sqrt{Q^2+U^2}/I=10^{-4}$, our assumed detection threshold. The scattered radiation is small compared to the stellar radiation. This is also the case for the dust thermal re-emission at $0.76\,\mu$m.  

These mass values are used to derive mass distributions for all of our disk models. These histograms are displayed in the top panel for each spectral type in Figure~\ref{belts}. There are 84 mass bins, or about 7.3 bins per mass decade. For each mass bin, we computed the mean value and a standard deviation for three relevant parameters in order to identify trends, if present. Keeping in mind that the size and definition of the computational grid is limited, we then analysed those trends to estimate which parameters have more influence on the polarization and the disk detectability. The mean values and their standard deviations for the parameters are displayed in Figure~\ref{belts}, below their respective histograms.

\subsubsection{Mass study \label{Mass}}

\begin{table} 
\begin{center}
\caption[Mass ranges.]{Mass ranges in M$_\oplus$ unit covered by our simulations for each belt and the four spectral types. The ranges are defined by their 10, 50 and 90 percentiles from the models. The masses correspond to the mass needed to produce a polarization of $10^{-4}$. }
\label{massranges}
\begin{tabular}{c|c|c|c|c}
\multicolumn{2}{c|}{}& Hot & Warm & Cold\\
 \hline
 \hline
A0 & 10\% & $1.0\times10^{-6}$ & $6.6\times10^{-4}$  & $1.6\times10^{-1}$ \\
& 50\% & $2.7\times10^{-5}$ & $2.5\times10^{-2}$ & $3.1$ \\
& 90\% & $1.6\times10^{-3}$ & $2.1$ & $4.1\times10^2$ \\
\hline
F5 & 10\% & $9.2\times10^{-8}$ & $4.6\times10^{-5}$ & $1.0\times10^{-2}$ \\
& 50\% & $1.9\times10^{-6}$ & $1.3\times10^{-3}$ & $1.9\times10^{-1}$ \\
& 90\% & $1.0\times10^{-4}$ & $1.1\times10^{-1}$ & $2.3\times10$ \\
\hline
K0 & 10\% & $1.2\times10^{-8}$ & $1.0\times10^{-5}$ & $1.5\times10^{-3}$ \\
& 50\% & $2.7\times10^{-7}$ & $1.7\times10^{-4}$ & $2.8\times10^{-2}$ \\
& 90\% & $1.7\times10^{-5}$ & $1.0\times10^{-2}$ & $3.0$ \\
\hline
M2 & 10\% & $9.6\times10^{-9}$ & $5.4\times10^{-6}$ & $1.1\times10^{-3}$ \\
& 50\% & $3.1\times10^{-7}$ & $1.6\times10^{-4}$ & $2.0\times10^{-2}$ \\
& 90\% & $2.2\times10^{-5}$ & $1.4\times10^{-2}$ & $2.5$ \\
\hline
\end{tabular}
\end{center}
\end{table}

The histograms in Figure~\ref{belts} show the presence of peaks for each type of belt. These peaks are the result of the grid discretization of the various parameters we used. They do not reflect an expected mass distribution for these belts. The values of the parameters selected are model dependent and represent best guesses for the range of values they might take. The choices made are based on models used to represent spectral energy distributions (SEDs) obtained from NIR to submm data currently available. To make progress, we are going to assume that the histograms give us a range of masses to be expected for the three types of belts. To make this simpler, we present in Table~\ref{massranges} the mass range covered by each belt according to their 10, 50 and 90 percentiles as determined from our simulations. 

With this assumption, we now perform comparisons between our model simulations and real measurements of belts in the literature to find out how feasible belt detection is with unresolved polarization. For hot belts around A0 stars, the literature typically obtains masses of $2\times 10^{-9}$ M$_\oplus$ \citep{2006A&A...452..237A,2008A&A...487.1041A,2011A&A...534A...5D,2017MNRAS.467.1614K} for Vega, $2.0\times 10^{-10}$ M$_\oplus$ for Fomalhaut \citep{2013A&A...555A.146L,2017MNRAS.467.1614K}, both of them much lower than the range of typical masses required to obtain a polarization of 10$^{-4}$ (Table~\ref{massranges}), starting at $1\times 10^{-6}$ M$_\oplus$. The G8 star $\tau$ Ceti has an exozodi disk first evaluated at $\sim 1\times 10^{-9}$ M$_\oplus$ by \cite{2007A&A...475..243D}. \cite{2017MNRAS.467.1614K} estimated a minimum mass of $3\times 10^{-11}$ M$_\oplus$ and a maximum mass from $1.2 \times 10^{-9}$ M$_\oplus$ (face-on) to $2.4\times 10^{-9}$ M$_\oplus$ (edge-on). $\tau$ Ceti can be compared to our F5 star. The median of the mass range corresponds to a hot disk of $2\times 10^{-6}$ M$_\oplus$ and the mass range goes down to almost $10^{-7}$ M$_\oplus$. \cite{2017MNRAS.467.1614K} modelled 6 other stars, including $\zeta$ Cep with a maximum mass up to $5\times 10^{-8}$ M$_\oplus$ and 10 Tau with a maximum mass up to $5\times 10^{-7}$ M$_\oplus$. Only the F9 IV-V star 10 Tau could, if the best conditions prevail, hope to be detected within our polarization threshold. Note that these two extreme values for $\zeta$ Cep and 10 Tau were not taken into account in Table \ref{tabpeak} as they are not typical values. As seen already with our toy model above (section~\ref{AA}), our own zodiacal cloud around our G2 star would not be detected according to our simulations if we observed it from a distance since it falls below the expected mass range. Since we assumed that those detections were the best case scenario, it implies that hot belts require a polarization precision better than $10^{-4}$, our detection threshold. 

Warm belts around solar-type stars have been studied more extensively by \cite{2009ApJ...705...89L} who covered a spectral range from F1 to K3. They used $Spitzer$ data and the models for their detected disks have dust masses from $\sim 4\times 10^{-8}$ M$_\oplus$ to $\sim 2\times 10^{-4}$ M$_\oplus$. These masses have been computed assuming 10$\,\mu$m grains only with a density for silicates of 3.3 g cm$^{-3}$. Comparing with results from the mass histograms which have masses ranging upwards from $5\times 10^{-6}$, $1\times 10^{-5}$, and $5\times 10^{-5}$ M$_\oplus$ for M2, K0 and F5 stars respectively, according to the 10 percentiles listed in Table~\ref{massranges}, we conclude that a few of the more massive warm dusty disks seen by $Spitzer$ (HD\,10647, HD\,38858 and HD\,45184) are slightly above our polarization detection threshold. The first of these stars also hosts a planet and a cold disk detected at 160$\,\mu$m by $Spitzer$ \citep{2009ApJ...704..109T}. The A star Fomalhaut has a warm disk at about 2 AU with a mass between 2 and 3 $\times 10^{-6}$ M$_\oplus$ \citep{2013A&A...555A.146L}, a value about a few times lower than the lower bound of the mass range for disks in our simulations. The more recent study by \cite{2016ApJ...826..171G} includes more stars and better models and the mass values for the whole sample are presented in Table \ref{tabpeak}. This paper reports 24 stars with $M > 1\times 10^{-5}$ M$_\oplus$, 16 stars with $M > 5\times 10^{-5}$ M$_\oplus$, 7 stars with $M > 1\times 10^{-4}$ M$_\oplus$ and 2 stars with $M > 1\times 10^{-3}$ M$_\oplus$. The stars with $M > 10^{-4}$ M$_\oplus$ are (with spectral types): HD\,15745 (F0), HD\,39060 (= $\beta$ Pic; A6\,V), HD\,80950 (A0\,V), HD\,106906 (F5\,V, a member of the Lower Centaurus Crux association), HD\,111520 (F5/6\,V), HD\,119718 (F5\,V) and HD\,145560 (F5\,V). These 7 stars have spectral types A or F and they all have a mass higher than the 10 percentiles listed in Table~\ref{massranges}.

For cold belts, lower bounds for detectable masses in polarimetry range from $1\times 10^{-3}$ M$_\oplus$ for M2 stars to about 0.2 M$_\oplus$ for A0 stars (Figure~\ref{belts} and Table~\ref{massranges}). The SONS JCMT/SCUBA-2 survey reports 46 detections of cold belts in a sample of 100 candidate stars \citep{2017MNRAS.470.3606H} with spectral types ranging from M2 to B8 and distances up to 96 pc (beyond the limit of our sample). Their dust masses computed from the submm fluxes range from $2\times 10^{-4}$M$_\oplus$ ($\tau$ Ceti) to 0.37 M$_\oplus$ (HD\,98800) with an average mass of 0.09 M$_\oplus$. The stars HD\,181327 \citep{2012A&A...539A..17L}, Fomalhaut \citep{2012A&A...540A.125A,2013A&A...555A.146L}, HD\,115617 and and HD\,207129 \citep{2009ApJ...704..109T} are included in the SONS detections. The stars HD\,10647 and HD\,38858 with detected warm disks mentioned above, were also detected by SONS. Therefore many cold disks should be detectable in polarimetry at approximately our detection threshold, if their configurations are suitable.

\subsubsection{Effect of the star's position}
\label{starpos}

\begin{figure}
\centering
\includegraphics[width=\columnwidth,trim = .5cm 0.5cm 2cm 0.5cm, clip]{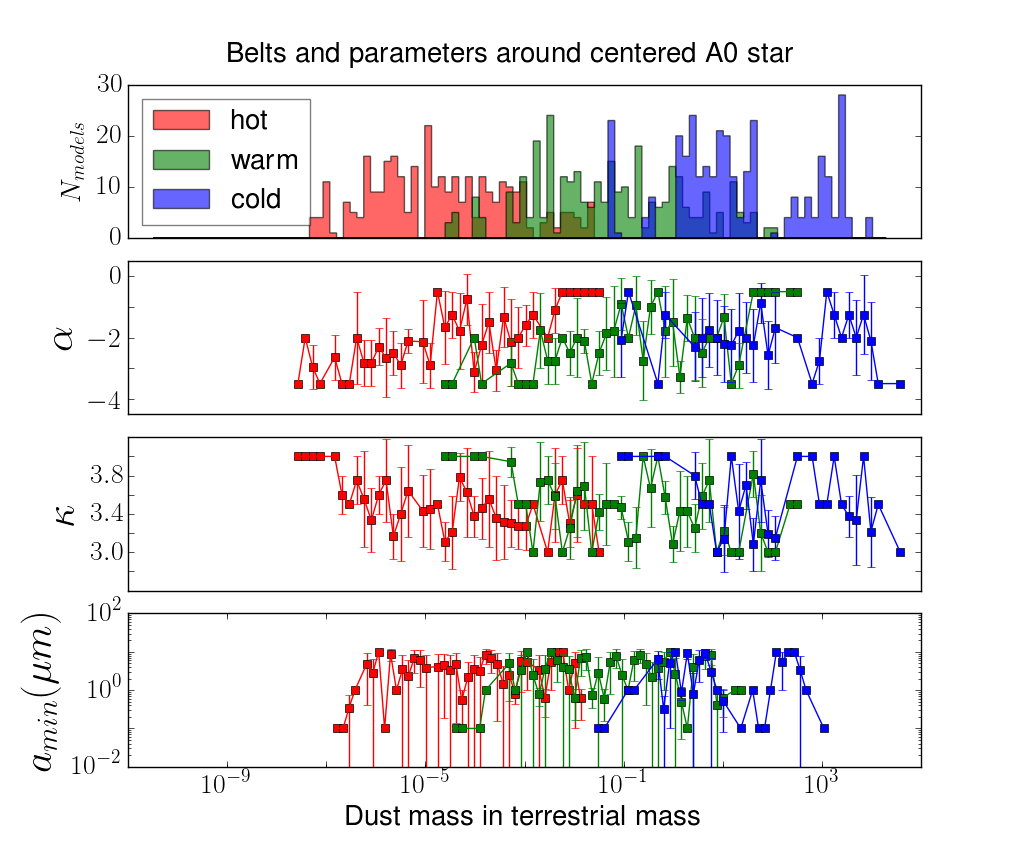}
\end{figure}

\begin{figure}
\centering
\includegraphics[width=\columnwidth,trim = .5cm 0.5cm 2cm 0.5cm, clip]{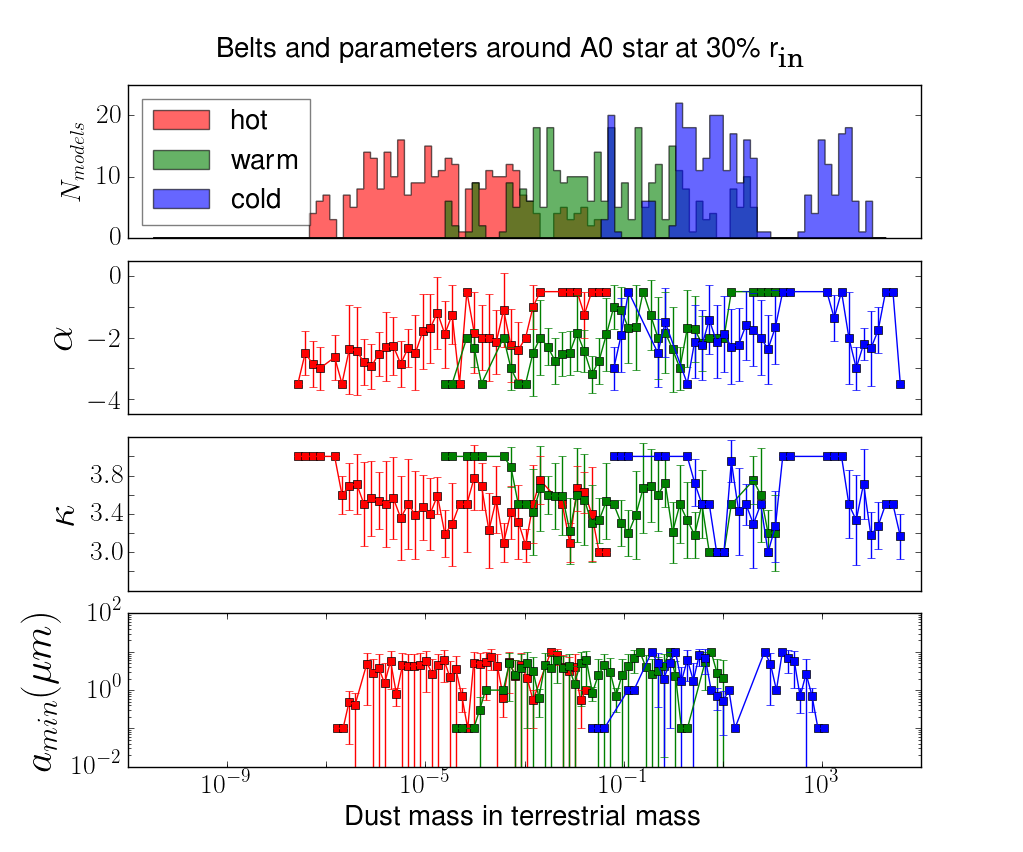}
\end{figure}

\begin{figure}
\centering
\includegraphics[width=\columnwidth,trim = .5cm 0.5cm 2cm 0.5cm, clip]{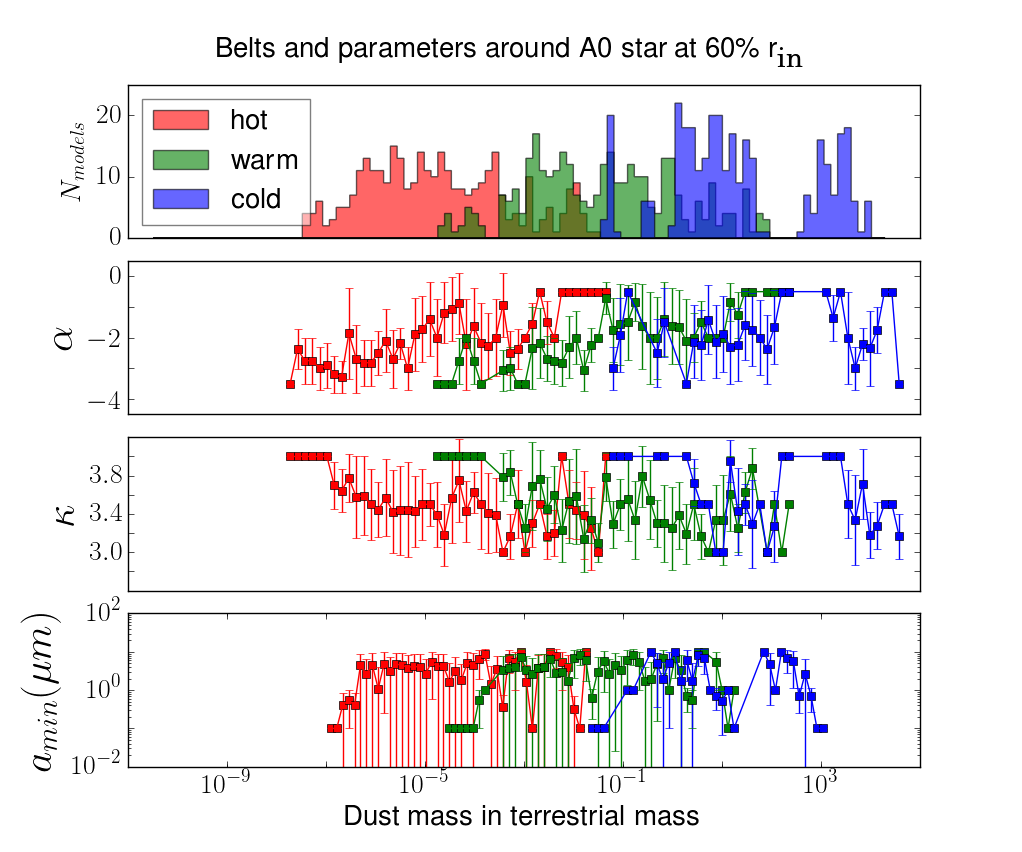}
  \caption{Histograms representing the number of simulations sorted by the mass of the minimum detectable mass of their disks. These histograms present results for only the A0 star and correspond to the top panel of Figure~\ref{belts} split into 3 parts to compare the outcome for each stellar position, centered, at 30 and 60 per cent of $r_{in}$. Underneath we plot the mean values in each mass bin for the surface density exponent $\alpha$, the grain-size distribution power-law index $\kappa$ and the minimum size of the grains, $a_{min}$, for the 3 stellar positions and 3 belts considered. The error bars correspond to 1 $\sigma$ and vary according to the number of simulations in each mass bin.}
  \label{belts2}
\end{figure}

As explained above, to study the polarization from elliptical disks, we moved the position of the star from 0 per cent to 30 per cent and 60 per cent of $r_{in}$. These results are shown in Figure~\ref{belts2} for the A0 star as an example. Note that the sum of the three panels of Figure~\ref{belts2} corresponds to the top panel of Figure~\ref{belts}. As before, the masses correspond to a polarization of $1\times 10^{-4}$.  
As we can see in Figure~\ref{belts2}, there are slight differences between the three simulations but smaller than one order of magnitude in the mass of the belts. This is confirmed by comparing 10, 50 and 90 percentiles for the three belts and the three positions of the A0 star in Table \ref{A0massranges} with each other and with those for the A0 star in Table \ref{massranges}. We also plotted cumulative distribution functions of the mass distributions for the belts considered in Table \ref{A0massranges} (not shown here). The curves for the three stellar positions are very similar, with only minor differences, confirming our conclusions.
Therefore, the stellar position is not as important as what one could expect. The explanation might be that part of the disk causes a larger polarization because of its proximity, but other parts produce a smaller polarization, compared to a circular disk; hence the global effect is modest only. We can conclude that position of the star has a second order effect on the polarization when we have circular disks. This conclusion probably applies also to eccentric disks. 

\begin{table} 
\begin{center}
\caption[Mass ranges A0 star.]{Mass ranges in M$_\oplus$ unit from our simulations for each belt and the three positions of the A0 star, centered, at 30 per cent and 60 per cent of $r_{in}$, as in Table \ref{massranges}. Note how the three stellar positions give comparable results. 
}
\label{A0massranges}
\begin{tabular}{c|c|c|c|c}
\multicolumn{2}{c|}{}& Hot & Warm & Cold\\
 \hline
 \hline
Centered & 10\% & $1.4\times10^{-6}$ & $9.0\times10^{-4}$  & $1.6\times10^{-1}$ \\
& 50\% & $2.9\times10^{-5}$ & $2.7\times10^{-2}$ & $3.5$ \\
& 90\% & $2.0\times10^{-3}$ & $2.4$ & $3.5\times10^2$ \\
\hline
30\%$r_{\rm in}$ & 10\% & $1.1\times10^{-6}$ & $6.8\times10^{-4}$ & $1.6\times10^{-1}$ \\
& 50\% & $2.8\times10^{-5}$ & $2.4\times10^{-2}$ & $3.4$ \\
& 90\% & $1.3\times10^{-3}$ & $1.8             $ & $4.4\times10^{+2}$ \\
\hline
60\%$r_{\rm in}$ & 10\% & $9.3\times10^{-7}$ & $5.9\times10^{-4}$ & $1.5\times10^{-1}$ \\
& 50\% & $2.3\times10^{-5}$ & $2.5\times10^{-2}$ & $2.5$ \\
& 90\% & $2.8\times10^{-3}$ & $2.1             $ & $4.0\times10^{+2} $ \\
\hline
\hline
\end{tabular}
\end{center}
\end{table}

\subsubsection{Parameter study}

In Figures~\ref{belts} and \ref{belts2} we plot three parameters we considered for this study to understand their relative importance on the detectability of disks. Unfortunately, we can only see rough tendencies and not fine effects because of the discretization of the parameters' values. This produces selection effects which we will try to determine in order to have contextualised results.

The effect of the surface density power-law index, $\alpha$\,\,$\epsilon$ \{$-3.5, -2.0, -0.5$\}, can be seen best for the hot and warm belts. As we could predict, sharper the disk is, smaller the mass needed will be. A sharp disk, i.e. with a high index implies that most of the mass will be closer to the star, which means more light to scatter for dust grains. This effect is not so clear for Kuiper belts since they are much farther from the star and so, the relative sprawl of the disk does not affect significantly the fraction of light received by the observer.

The minimum dust grain size, $a_{min}$, appears to be a critical element in the analysis: it confirms what we expected and is consistent with what was obtained with our toy model with only a single grain size (section \ref{AA}): when the minimum size diminishes, the mass needed for polarization detection is smaller. It is due to the higher relative cross-section of tinier bodies compared to their mass. Therefore, when more tiny dust grains are available, more scattered, and polarized, light will be produced, implying an easier detection. But we tested only three values ($10^{-1}, 1, 10$) $\mu$m for this parameter. So, when we reach high masses, we notice a saturation effect. This is an effect due to the selection which is probably not physical: by introducing even bigger sizes, we might have seen that the plot continues to grow.

The size distribution power-law exponent, $\kappa$, acts the same way as the minimum size: it controls the number of tiny dust grains. The higher the exponent is, the tinier the grains will be (on average). So, obviously, the lowest detectable masses are those with the highest exponent which makes the size distribution sharper.

\section{Discussion}
\label{Discussion}

\subsection{Interstellar polarization in the solar neighbourhood}

Up until now, we purposely analysed the sample of stars as if their polarization is due entirely to debris disks. The fact that only two stars detected in polarization are associated with debris disks detected by mid- and far-IR excess (see section \ref{sub:3sigma} and Table \ref{tab:3sigma}) suggests that something else is contributing to the polarization detected in these stars. Indeed, even if our sample stars are nearby, it is possible or likely that the detected polarization is due to interstellar dust aligned by magnetic fields. \cite{1993A&A...274..203L} used his polarization catalogue to confirm the previous result by \cite{1982A&A...105...53T} about a significant depletion of dust within 35 pc from the Sun. \cite{1999A&A...346..955L} updated his results when the $Hipparcos$ distances became available and showed that previous distances had been underestimated. As a result, the  cavity walls of our local bubble are located at 70 to 150 pc, varying with direction. In addition to not being spherical, the local  bubble is hot, as X-rays have been detected \citep{1990ApJ...354..211S,2011ARA&A..49..237F}. This means that our sample stars are all located $inside$ the very low-density local bubble since their distances vary from up to 9 pc for M-type stars to up to 46 pc for A-type stars.

More recent observations with high precision polarimeters confirm the early observations. \cite{2010yCat..74052570B} and \cite{2016MNRAS.455.1607C,2016MNRAS.460...18C} presented surveys of nearby bright stars in the northern and southern hemispheres, respectively. They show that even in the low-density local bubble, the polarization increases linearly with distance as $P/d \sim 1 \times 10^{-6} $ pc$^{-1}$ \citep{2016MNRAS.455.1607C,2016MNRAS.460...18C}, instead of $2 \times 10^{-5}$ pc$^{-1}$ in the general interstellar medium \citep{1959VeGoe...7..200B} outside the local bubble. In the direction of the North Galactic pole, PlanetPol measured a value as low as $P/d \sim 2 \times 10^{-7} $ pc$^{-1}$ \citep{2010yCat..74052570B} inside our local bubble. Stars with a polarization compatible with this linear distance dependence ($\sim 1 \times 10^{-6} $ pc$^{-1}$) and  aligned with other stars in similar directions on the sky would have a polarization compatible with interstellar polarization. Stars with different (larger) polarizations than given by this relation and orientation different than their neighbours' can be suspected of having a component of their polarization due to a debris disk. 

The two stars with debris disks and detected polarization, HD\,7570 and HD\,165908, both have a linear polarization much higher than what these relations predict for a distance of 15 pc, $P/d \sim  49.6 \pm 4.0 \times 10^{-6}$ pc$^{-1}$ and $9.6 \pm 3.2 \times 10^{-6}$ pc$^{-1}$ respectively, and therefore cannot be explained by local interstellar polarization. 

In Table \ref{tab:3sigma} we present 16 other systems detected in polarization and without a FIR excess except one with an unknown status. We also computed their $P/d$ ratio and all of them have a significant ratio, i.e. $P/d > 3 \sigma$ and also $P/d > 1 \times 10^{-6}$ pc$^{-1}$, the expected linear polarization dependence within our local bubble. Therefore all of them are possible candidates to have an intrinsic polarization component which could be due to a debris disk not detected by their FIR excess yet (see also section \ref{Hprec-polar}). However, we caution that twelve of those 16 stars (not including HD\,165908) have a detected polarization $3.0 < P/\sigma_P < 4.0$ and should be confirmed with high precision polarimeters. 

\subsection{Cold disks are more favourable for polarization detection}
\label{Cold-disks}

In section \ref{Mass} above, we compared the model mass required to obtain a polarization equal to a polarization threshold of $10^{-4}$ to observed masses  for the three different belt types and a range of typical spectral types. We found that observed masses of hot disks fall significantly (by factors of 100 to 1000) below masses corresponding to  the polarization threshold; the mass a few warm disks barely reaches the masses of the threshold. However, many observed cold disks have masses above the masses required for the polarization threshold. The polarization threshold was chosen to correspond approximately to the limit that $traditional$ polarimeters can reach  in routine observations, although they can with great care exceed it somewhat (see, e.g., Table \ref{tab:3sigma}). It is now clear with the above comparison that essentially only the massive cold disks have a good chance to be detected with traditional polarimeters. 

Naively, one might have expected to measure the polarization more easily for hot belts than for cold ones, since dust grains located in hot belts present a larger solid angle as seen from the star than dust grains farther out in cold disks because of their proximity to their star. Effectively, simulations show that smaller masses are required to obtain a polarization of $10^{-4}$ (Table \ref{massranges}) for hot disks than for warm and cold ones. But these results do not match with observations. Observed masses, as determined by fits to SEDs and resolved imaging data, are too small for hot disks, barely reach required masses for warm disks and are just about right for cold disks to be detected with unresolved polarimetry in the visible.

One way to explain this paradox is to use the two-parameter model of debris disks presented by \cite{2008ARA&A..46..339W}. Such a simplification is possible because the SED of known debris disks can be represented reasonably well by a black body with a single temperature, $T$. The other parameter is the fractional luminosity, defined as the ratio of the IR luminosity of the dust to that of the star, $f = L_{IR}/L_*$. In his model, \cite{2008ARA&A..46..339W} shows that the disk mass expressed in M$_\oplus$ is a scaled version of the fractional luminosity,
\begin{equation}
M_{disk}=12.6f r^2 \kappa_{\nu}^{-1} X_{\lambda}^{-1},
\label{Mdisk}
\end{equation}
where $r$ is the disk radius in AU, the opacity $\kappa_{\nu}=$ 45 AU$^2$ M$_\oplus^{-1}$ (= 1.7 cm$^2$ g$^{-1}$) at 850~$\mu$m, and $X_{\lambda}=1$ for $\lambda < 210$ ~$\mu$m and $X_{\lambda}={\lambda}/210$ otherwise. \cite{2007ApJ...660.1556R} have shown that $M_{disk}/f \propto r^2$ holds across the FIR and submm regimes. Since we know that disk mass is proportional to polarization (cf. equation \ref{Ptot-eq}), we get:
\begin{equation}
P \propto f r^2.
\label{Pr-rel}
\end{equation}
This equation shows that cold disks located further away than hot and warm disks should have a higher polarization. Also, disks with a higher fractional luminosity $f$ are the best ones to observe as they should have a larger polarization. In fact,  the fractional luminosity defines the total cross-sectional area of optically thin disks \citep{2008ARA&A..46..339W}: $\sigma_{tot}=4 \pi r^2 f$, where $\sigma_{tot}$ is in AU$^2$. Therefore, cold disks are more efficient at polarizing stellar light simply because they have more dust grains, hence more mass, than closer disks. 

As pointed out by \cite{2010RAA....10..383K}, this simple model is not perfect since  mass and polarization do not have the same dependence on the grain size distribution, $n(a)$. Mass hides mostly where $n(a) a^3$ is maximum and is affected by $a_{max}$, the maximum grain size,  whereas polarization peaks approximately where $x = k a = 2 \pi a / \lambda \approx 2$, where $\lambda$ is the wavelength of observation. For our 0.76~$\mu$m  data, this gives $a \approx 0.2$ -- $0.3$~$\mu$m, significantly smaller than grains contributing most of the detected mass, with $a \sim 1~$mm (e.g. \cite{2017MNRAS.470.3606H}). Despite this difference in the most sensitive grain radii for mass and polarization, the model does offer a rough understanding of the physics involved and helps understanding why cold disks have larger polarizations.

\subsection{High precision polarimetry and polarization detection}
\label{Hprec-polar}
High precision polarimeters have begun to observe stars with debris disks \citep{2010yCat..74052570B,2016MNRAS.455.1607C,2016MNRAS.460...18C,2016ApJ...825..124M} and indeed can measure their polarization. So far, 15 stars with known debris disks and two with IR excess have been observed with high precision polarimeters and the data published. \cite{2010yCat..74052570B} observed 6 of them and discussed the results for Vega in detail. Its polarization is 17.2 $\pm$ 1.0 ppm ($10^{-6}$). However its debris disk extends from 11\,arcsec to 105\,arcsec according to $Spitzer$ observations \citep{2005ApJ...628..487S} and was therefore outside the aperture of PlanetPol. There is also evidence for circumstellar material within 1\,arcsec, probably a hot disk. However the disk is seen essentially face-on ($i = 4.7\degr$) hence the authors conclude that its polarization is compatible with being entirely of interstellar origin. Their five other debris disk stars are discussed by \cite{2016MNRAS.455.1607C,2016MNRAS.460...18C}. Merak ($\beta$ UMa) and  $\beta$ Leo are compatible with an interstellar origin and probably also $\alpha$ CrB but this last one is also an Algol-type eclipsing binary. The best case is $\gamma$ Oph which has a polarization of 40 ppm but given its distance of 29 pc, its polarization is compatible with an interstellar origin. However its polarization angle is aligned with the minor axis of the imaged disk \citep{2016MNRAS.455.1607C,2016MNRAS.460...18C}, the expected polarization orientation. It is therefore likely that $\gamma$ Oph has two polarization components, intrinsic and interstellar. \cite{2016ApJ...825..124M} observed 6 stars with strong excess at NIR wavelengths, an indicator of hot dust belts, and did not detect the expected strong polarization in any of them. They ruled out scattered light as the origin of their emission as the probability of all six of them being seen face-on is small. The results of our simulations presented above show that the level of polarization expected for hot belts are difficult to observe, even with high precision polarimeters. 
 
With the performances of high precision polarimeters, we can certainly detect some debris disks in polarimetry. If a polarization level of $10^{-4}$ was high enough to reject easily or neglect the interstellar component of the polarization in the past, this is not the case for polarizations as low as $10^{-6}$. \cite{2016MNRAS.455.1607C,2016MNRAS.460...18C} analysed the interstellar polarization inside the solar bubble and concluded that, depending on the direction observed, the polarization increase with distance is between $2\times10^{-7}$ and $2\times10^{-6}$ pc$^{-1}$. The lowest rate found in our detected measurements in Table \ref{tab:3sigma} is $(5 \pm 1) \times10^{-6}$ pc$^{-1}$. So, all the polarizations detected are above the level of interstellar polarization in the solar bubble by at least a factor of 2 and are therefore likely to have an intrinsic polarization component. With the lower polarization levels reached by the new polarimeters, disentangling interstellar and intrinsic debris-disk polarizations is now a real issue to be considered. This is particularly the case for stars located near the galactic plane. 
One way to address this issue is to take measurements in different filters to obtain the wavelength dependence of the polarization and compare with the expected dependence of the interstellar polarization. If there are significant deviations from Serkowski's law, and in particular if the observed polarization angles rotate as a function of wavelength, then a least-squares fit method can be used to separate the two components, as done by \cite{1979AJ.....84..812P} for classical Be stars. Such a method works if the two polarization components, intrinsic and interstellar, are not collinear in the Stokes $Q$-$U$ plane, and if the data are of sufficient quality. Since this method will be more time consuming, it will be appropriate for studying a few stars at a time, not for a large survey. 

We showed in section \ref{analysis} that there is no statistically significant difference between samples with debris disks and those without, according to KS tests and their cumulative distribution functions. \cite{2015RMxAA..51....3G} observed 88 southern hemisphere stars, with and without mid-IR excess based on $Spitzer$ observations and with aperture polarimetry in 4 different filters. They combined their sample with an earlier version of ours, \cite{2010Simon}, selecting stars with and without known debris disks, based on IR excess emission, and removing stars with unknown status. With their 51 IR-excess stars and 97 stars without IR excess, they found similar results with their CDF and statistical analyses than ours.

\section{Conclusion and Further Prospects}

We performed a coherent census of polarization due to nearby debris disks for 109 stars. The stars were selected from the DEBRIS and DUNES candidate stars observed with $Herschel$. Combining with polarization measurements from the literature for other candidate stars, we obtained a list of 223 stars with also information about the presence of debris disks based on their mid- and far-IR excesses. Eighteen of them were detected with a polarization $P \ge 3 \sigma_P$. We found that the polarization distribution of the samples with and without debris disks are not statistically different. Among the eighteen stars with detected polarization, two of them have a debris disk according to their IR excesses. One of them, HD\,165908 is a binary with an imaged disk; its polarization is parallel to the pericentre direction, within a few degrees, as expected. The other star, HD\,7570, is single and has not been imaged yet, hence we predict the orientation of its disk, position angle $\approx 138\degr$, perpendicular to its strong polarization. 

There are many factors which can explain the low-detection rate result, separately or in combination:\\
1. Only about 20 per cent of stars have one or more debris disks according to their mid-IR and FIR excesses \citep{2013A&A...555A..11E,2015A&A...582L...5R,2016A&A...593A..51M};\\
2. The dust mass present may not be sufficient to produce a detectable polarization, or equivalently the fractional luminosity of the disk is too small (see sections \ref{Numresults} and \ref{Cold-disks});\\
3. The inclination of the disk may be such that the polarization as seen by the observer cancels out (mostly face-on disks);\\
4. Disks around some stars are too large to fit within the aperture used by polarimeters (e.g., the case of $\alpha$ Lyr discussed above);\\
5. Dust grains may not be good scatterers in visible bands, such as nano-scale size dust grains trapped in hot belts \citep{2016ApJ...825..124M,2015RMxAA..51....3G,2016ApJ...816...50R}.

Factor 2 above (insufficient dust) can be mostly overcome with high precision polarimeters. One can use a larger diaphragm to solve factor 4, to some extent, as long as there is no background star within the larger aperture. As discussed above, interstellar polarization needs to be accounted for in the analysis of the polarization data.

The analytical model and numerical simulations are consistent with observational results. We computed mass histograms corresponding to a polarization threshold of $10^{-4}$ for cold, warm and hot disks with a large grid of model parameters. Comparison with masses obtained from observations of IR excesses shows that simulated hot belts yield masses larger than the observed ones by up to a factor of 1000. The masses of only a few warm disks observed by $Spitzer$ reach those of models corresponding to the polarization of $10^{-4}$. However, this polarization level produces disk masses reached and exceeded by many cold disks as seen by the JCMT/SCUBA-2 SONS survey \citep{2017MNRAS.470.3606H}. These simulations show that cold disks can be detected by traditional polarimeters, but high precision polarimeters are needed for detection of warm and hot disks. Of course, the caveats or factors mentioned above apply.

This result can be explained by the fact that polarization $P \propto f r^2 \propto \sigma_{tot}$, the total cross-sectional area of dust grains in optically thin disks, as derived from the two-parameter debris disk model (section \ref{Cold-disks}). Cold disks have a larger polarization because they have more dust grains and more mass than closer disks. 

The simulations also showed that eccentric disks have only minor effects on the level of polarization, at least as much as represented by moving the position of the star interior to circular disks. The effect of the slope of the surface density distribution is better seen in hot and warm disks. The larger the slope, the smaller the mass needed to produce a given polarization. With the limitation of only three minimum grain sizes, more scattered and polarized light is produced when the minimum grain size is reduced. This can be explained by the higher relative cross sections of small grains for a given mass. Finally, the slope of the size distribution acts in the same manner: more small grains means higher polarization.

\section*{Acknowledgements}
The authors wish to sincerely thank the telescope operators at the Mont-M\'egantic Observatory, Pierre-Luc L\'evesque, Bernard Malenfant and Ghislain Turcotte. We also would like to thank R\'emi Fahed, Patrick Ingraham and Lison Malo for their precious help, Marie-Mich\`ele Limoges for a careful reading of a previous version of this manuscript and David Lafreni\`ere for insightful discussions about HR 8799. We thank the anonymous referee whose constructive comments and requests have helped improve significantly  the paper. This research was supported in part by a grant from the Conseil de Recherche en Sciences Naturelles et en G\'enie du Canada. J.-C.\,A. acknowledges support from the Programme National de Plan\'etologie (PNP) of CNRS/INSU co-funded by the CNES. This research has made use of the SIMBAD database and the VizieR catalog access tool, operated at CDS, Strasbourg, France.
  
\nocite{*} 
  
\bibliography{biblio.bib}


\pagebreak
\appendix
\section{All stars from the sample with a polarization $P < 3\sigma_P$}
\label{Appen}

We present in Table~\ref{tablestars} below stars in our sample with a polarization $P < 3\sigma_P$. Stars with a detected polarization, $P \ge 3 \sigma_P$, are given in Table~\ref{tab:3sigma}. The polarization data for the sample stars come from OMM and from the Leroy compilation. The description of these two tables is given in section~\ref{StatComp}.

\begin{table*}		
\begin{ThreePartTable}
\caption{Polarization data for all the stars observed at OMM and from the Leroy compilation with a polarization $P < 3 \sigma_P$.} 		
\label{tablestars}
\begin{center}
\begin{onecolumn}
\begin{tabular}{clrrrcllcrl}
\hline
UNS  & name & $P$ & $\sigma_P$ & $\theta$(\degr) & $\sigma_\theta$(\degr) & 
$P/\sigma_P$ & Distance &FIR\tnote{a} &Observer\tnote{b}& Date\tnote{c}\\
ID &&($10^{-5}$)&($10^{-5}$)&&&&(pc)&excess&&\\
\hline\hline
M003&HD 95735&47&31&177&23&1.5&2.543$\pm$0.001&N&SI&02 Mar\\
M015&HIP 36208&11&30&26&34&0.4&3.795$\pm$0.018&?&SI&02 Mar\\
M031&HIP 54211&32&35&ind.\tnote{d}&52&0.9&4.862$\pm$0.022&?&SI&02 Mar\\
M032&GJ 388&78&50&54&15.&1.6&4.888$\pm$0.067&?&SI&23 Jan\\
M040&HD 119850&116&93&79.&20&1.2&5.395$\pm$0.030&N&SI&22 Jan\\
M042&HD 265866&0&36&ind.&52&0.0&5.614$\pm$0.04&N&SI&21 Jan\\
M053&HIP 37766&0&31&130&44&0.0&5.982$\pm$0.073&?&SI&02 Ma.\\
M054&HIP 34603&0&46&ind.&52&0.0&6.119$\pm$0.067&?&SI&22 Jan\\
M060&GJ 661 A&54&44&144&20&1.2&6.397$\pm$0.052&?&SI&02 Mar\\
M067&HIP 53767&0&30&ind.&52&0.0&6.697$\pm$0.071&?&SI&02 Mar\\
M069&GJ 3522&103&36&79.&11.&2.8&6.772$\pm$0.09&?&SI& 02 Mar\\
M070&HIP 53020&54&41&75&18&1.3&6.794$\pm$0.137&?&SI&02 Mar\\
M076&HIP 51317&20&37&144&23&0.5&7.129$\pm$0.103&?&SI&02 Mar\\
M090&GJ 1093&170&120&57&19&1.4&7.764$\pm$0.211&?&SI&22 Jan\\
M095&HIP 49986&56&34&152&22&1.6&7.930$\pm$0.114&?&SI&02 Mar\\
M100&HIP 86287&0&33&ind.&52&0.0&8.050$\pm$0.097&?&SI&02 Mar\\
M110&GJ 1230 A&83&50&ind.&ind.&1.7&8.271$\pm$0.493&?&SI&02 Mar\\
M109&HIP 38956&31&45&ind.&52&0.7&8.269$\pm$0.159&?&SI&22 Jan\\
K005&HD 88230&82&48&160&19&1.7&4.866$\pm$0.012&?&SI&22 Jan\\
K011&HD 79210&14&40&118&22&0.4&6.108$\pm$0.094&N&SI&08 Dec\\
K014&HD 16160&14&13&ind.&ind.&1.1&7.191$\pm$0.023&N&LE\\
K016&HD 4628&32&11&ind.&ind.&2.9&7.449$\pm$0.027&N&SC LE\\
K017&HD 10476&16&6&166&19&2.7&7.533$\pm$0.028&N&SI SE MA&23 Jan\\
K019&HD 216803&7&18&ind.&ind.&0.4&7.611$\pm$0.036&N&SC\\
K021&HD 157881&33&33&ind.&52&1.0&7.700$\pm$0.042&?&SI&02 Mar\\
K027&HD 192310&10&13&ind.&ind.&0.8&8.910$\pm$0.024&N&LE\\
K028&HD 103095&9&37&177&21&0.2&9.081$\pm$0.033&?&SI AP&22 Jan\\
K031&HD 151288&59&35&81&18&1.7&9.809$\pm$0.067&N&SI&02 Mar\\
K041&HIP 66459&66&36&96&12&1.8&10.935$\pm$0.135 &?&SI&02 Mar\\
K055&15009+4526A&65&38&125&15.&1.7&11.881$\pm$0.147&?&SI&02 Mar\\
K056&HD 97101&53&24&135.&10.&2.2&11.961$\pm$0.120&?&SI&04 Dec\\
K060&HD 75732&99&42&136.&11.&2.4&12.460$\pm$0.104&?&SI& 21 Jan\\
K064&HD 82106&31&32&ind.&52&1.0&12.894$\pm$0.106&?&SI&02 Mar\\
K070&HIP 67090&52&41&57&16&1.3&13.193$\pm$0.169&?&SI&22 Jan\\
K072&HD 128165&0&37&175&48&0.0&13.215$\pm$0.073&?&SI&02 Mar\\
K074&HD 120476&12&33&ind.&52&0.4&13.373$\pm$0.136&?&SI&02 Mar\\
K082&HIP 27188&0&30&ind.&52&0.0&13.716$\pm$0.176&?&SI&21 Jan\\
K086&GJ 400 A&70&29&132.&13.&2.4&13.900$\pm$0.316&?&SI& 04 Dec\\
K092&HD 110315&98&49&88.&14.&2.0&14.194$\pm$0.146&?&SI& 22 Jan\\
K096&HIP 70218&35&36&ind.&52&1.0&14.393$\pm$0.169&?&SI&02 Mar\\
K098&HD 144579&0&46&ind.&52&0.0&14.508$\pm$0.069&?&SI&02 Mar\\
K099&HIP 37288&0&31&ind.&52&0.0&14.533$\pm$0.292&?&SI&02 Mar\\
K108&HIP 13375&0&64&ind.&52&0.0&14.757$\pm$0.299&?&SI&23 Jan\\
K111&HD 110833&0&33&ind.&52&0.0&14.889$\pm$0.146&?&SI&02 Mar\\
K121&GJ 319 A&0&29&105&28&0.0&15.368$\pm$0.399 &?&SI&02 Mar\\
G005&HD 20794&3&6&ind.&ind.&0.5&6.043$\pm$0.007&Y&SC TI KO MA\\
G006&HD 131156&8&8&ind.&ind.&1.0&6.708$\pm$0.021&N&LE HU\\
G007&HD 109358&15&7&67&34&2.1&8.440$\pm$0.014&N&SI PI TI BE&02 Mar\\
G008&HD 115617&10&6&ind.&ind.&1.7&8.555$\pm$0.016&Y&SE TI LE MA\\
G010&HD 114710&2&4&ind.&ind.&0.5&9.132$\pm$0.014&N&PI TI SE HU\\
G011&HD 20630&6&6&ind.&ind.&1.0&9.144$\pm$0.022&N&SC TI HU LE\\
G012&HD 102365&21&8&ind.&ind.&2.6&9.221$\pm$0.020&N&TI MA\\
G016&HD 13974&17&7&51&24&2.4&10.778$\pm$0.045&N&SI TI PI BE&23 Jan\\
G017&HD 82885&17&8&ind.&ind.&2.1&11.363$\pm$0.041&N&PI LE BE\\
G019&HD 141004&1&7&ind.&ind.&0.1&12.122$\pm$0.045&N&TI LE BE\\
G025&HD 133640&20&31&ind.&52&0.6&12.740$\pm$0.097&N&SI&02 Mar\\
G026&HD 10307&13&10&ind.&ind.&1.3&12.739$\pm$0.044&N&PI TI BE&\\
G029&HD 30495&7&11&129&18&0.6&13.273$\pm$0.063 &N&SI LE&08 Dec\\
G033&HD 95128&35&12&ind.&52&2.9&14.062$\pm$0.049&?&SI TI&04 Dec\\
G040&HD 86728&7&11&ind.&52&0.6&15.052$\pm$0.071&?&SI LE&22 Jan\\
\end{tabular}
\end{onecolumn}
\end{center}
\end{ThreePartTable}
\end{table*}

\begin{table*}		
\begin{ThreePartTable}
\contcaption{Polarization data for all the stars observed at OMM and from the Leroy compilation with a polarization $P < 3\sigma_P$.}
\begin{center}
\begin{onecolumn}
\begin{tabular}{clrrrcllcrl}
\hline
UNS  & name & $P$ & $\sigma_P$ & $\theta$(\degr) & $\sigma_\theta$(\degr) & 
$P/\sigma_P$ & Distance &FIR\tnote{a} &Observer\tnote{b}& Date\tnote{c}\\
ID &&($10^{-5}$)&($10^{-5}$)&&&&(pc)&excess&&\\
\hline\hline
G057&HD 111395&18&23&18&20&0.8&16.938$\pm$0.129&?&SI AP&23 Jan\\
G061&HD 122742&29&10&38&13.&2.9&16.965$\pm$0.178&?&SI LE&22 Jan\\
G065&HD 50692&5&13&54&19&0.4&17.237$\pm$0.122&?&SI LE&21 Jan\\
G067&HD 142267&20&30&ind.&ind.&0.7&17.349$\pm$0.163&N&MA\\
G079&HD 99491&9&16&ind.&ind.&0.6&17.777$\pm$0.210&N&LE\\
G081&HD 137108&12&35&ind.&52&0.3&17.923$\pm$0.251&?&SI&02 Mar\\
G086&HD 84737&13&11&106&15.&1.2&18.344$\pm$0.094&N&SI LE&04 Dec, 22 Jan\\
G087&HD 222335&28&24&ind.&ind.&1.2&18.578$\pm$0.217 &N&SC\\
G088&HD 154345&0&55&ind.&52&0.0&18.582$\pm$0.110&?&SI&02 Mar\\
G089&HD 4747&39&22&ind.&ind.&1.8&18.672$\pm$0.188 &N&SC\\
G092&HD 9540&30&19&ind.&ind.&1.6&19.034$\pm$0.166&N&SC\\
G093&HD 52711&11&13&176&17.&0.8&19.163$\pm$0.150&?&SI LE&21 Jan\\
G094&HD 78366&166&75&150.&14.&2.2&19.183$\pm$0.121&?&SI&21 Jan\\
G096&HD 43587&23&14&ind.&ind.&1.6&19.250$\pm$0.148&N&LE\\
G100&HD 79028&20&25&ind.&52&0.8&19.569$\pm$0.122&?&SI LE&22 Jan\\
G101&HD 136923&29&30&ind.&52&1.0&19.600$\pm$0.242&N&SI&02 Mar\\
G107&HD 212698&15&11&ind.&ind.&1.4&20.191$\pm$0.487&N&SC\\
G108&HD 89269&46&52&177&22&0.9&20.243$\pm$0.205&N&SI&22 Jan\\
G109&GJ 337 A&118&61&118&17.&1.9&20.369$\pm$0.227 &?&SI&23 Jan\\
G113&HD 212330&4&8&ind.&ind.&0.5&20.569$\pm$0.144&N&SC\\
G117&HD 197076&23&15&ind.&ind.&1.5&20.896$\pm$0.206 &N&LE\\
G118&HD 1835&11&7&ind.&ind.&1.6&20.923$\pm$0.229&N&LE SC\\
G121&HD 117043&27&14&ind.&ind.&1.9&21.160$\pm$0.139&N&LE\\
G122&HD 146361&7&8&ind.&ind.&0.9&21.196$\pm$0.485&N&LE\\
F001&HD 61421&5&8&ind.&ind.&0.6&3.507$\pm$0.013&?&TI SE SP\\
F003&HD 30652&13&9&ind.&ind.&1.4&8.069$\pm$0.011&N&PI TI BE\\
F004&HD 98231&70&32&28.&15.&2.2&8.368$\pm$0.055&N&SI&04 Dec\\
F005&HD 1581&6&5&ind.&ind.&1.2&8.586$\pm$0.012&N&TI MA\\
F006&HD 38393&2&6&ind.&ind.&0.3&8.926$\pm$0.014&Y&LE SE TI MA+\\
F007&HD 203608&15&7&ind.&ind.&2.1&9.261$\pm$0.016&N&TI\\
F011&HD 142860&13&7&ind.&ind.&1.9&11.255$\pm$0.023&N&TI MA LE\\
F012&HD 33262&14&8&ind.&ind.&1.8&11.645$\pm$0.024&Y&TI\\
F014&HD 110379&23&8&ind.&ind.&2.9&11.745$\pm$0.080&N&TI BE\\
F015&HD 207098&14&6&ind.&ind.&2.3&11.869$\pm$0.02&N&WA TI\\
F018&HD 90839&9&7&ind.&52&1.3&12.785$\pm$0.047&?&SI PI TI+ BE&04 Dec\\
F019&HD 82328&2&6&ind.&52&0.3&13.481$\pm$0.024&N&SI PI TI BE LE&22 Jan\\
F022&HD 22484&20&7&10&24&2.9&13.977$\pm$0.105&Y&SI TI&23 Jan\\
F023&HD 20010&6&9&&&0.7&14.235$\pm$0.091&N&SC\\
F024&HD 17206&11&5&170&14.&2.2&14.237$\pm$0.365&N&SI SC TI BM&23 Jan\\
F026&HD 126660&9&12&ind.&ind.&0.8&14.528$\pm$0.030&N&PI TI BE\\
F027&HD 197692&17&7&ind.&ind.&2.4&14.677$\pm$0.058&N&TI LE MA\\
F030&HD 105452&12&8&ind.&ind.&1.5&14.936$\pm$0.036&N&TI\\
F036&HD 120136&79&38&74.&15.&2.1&15.622$\pm$0.049&?&SI& 22 Jan\\
F039&HD 128167&14&7&ind.&ind.&2.0&15.828$\pm$0.065&N&PI TI BE BM\\
F043&HD 215648&8&4&ind.&ind.&2.0&16.296$\pm$0.053&N&SC PI BE MA\\
F044&HD 48682&5&10&139&29&0.5&16.714$\pm$0.084&?&SI LE&04 Dec\\
F045&HD 55575&9&10&161&22&0.9&16.889$\pm$0.094&?&SI LE&22 Jan\\
F046&HD 17051&7&7&ind.&ind.&1.0&17.168$\pm$0.065 &N&KO SC MA\\
F048&HD 81997&20&30&ind.&ind.&0.7&17.313$\pm$0.602 &N&MA\\
F053&HD 23754&6&7&ind.&ind.&0.9&17.609$\pm$0.059&N&TI\\
F055&HD 114378&24&13&48&32&1.8&17.828$\pm$0.282 &N&SI LE HU AP&22 Jan\\
F058&HD 58946&18&7&11&19&2.6&18.022$\pm$0.078&N&SI PI TI BE&22 Jan\\
    &GJ274   &23&35&ind.&52&0.6&18.047$\pm$0.039&?&SI&08 Dec\\
F061&HD 69897&2&14&126&21&0.1&18.268$\pm$0.107&?&SI LE&21, 22, 23 Jan\\
F062&HD 129502&11&7&ind.&ind.&1.6&18.282$\pm$0.067&N&TI LE MA\\
F063&HD 109085&11&16&ind.&ind.&0.7&18.282$\pm$0.060&Y&LE BM\\
F066&HD 202275&8&6&ind.&ind.&1.3&18.218$\pm$0.083&N&WA TI LE\\
F067&HD 56986&46&16&147&15.&2.9&18.515$\pm$0.228&N&SI BE&21, 22 Jan\\
F084&GJ 335 B&54&54&171&37&1.0&20.378$\pm$0.153&N&SI&22 Jan\\
    &HD 78154&33&12&ind.&ind.&2.7&520.378$\pm$0.153&N&TI&\\
F085&HD 27290&22&8&ind.&ind.&2.8&20.461$\pm$0.151&Y&TI\\

\end{tabular}
\end{onecolumn}
\end{center}
\end{ThreePartTable}
\end{table*}

\begin{table*}		
\begin{ThreePartTable}
\contcaption{Polarization data for all the stars observed at OMM and from the Leroy compilation with a polarization $P < 3\sigma_P$.}
\begin{center}
\begin{onecolumn}
\begin{tabular}{clrrrcllcrl}
\hline
UNS  & name & $P$ & $\sigma_P$ & $\theta$(\degr) & $\sigma_\theta$(\degr) & 
$P/\sigma_P$ & Distance &FIR\tnote{a} &Observer\tnote{b}& Date\tnote{c}\\
ID &&($10^{-5}$)&($10^{-5}$)&&&&(pc)&excess&&\\
\hline\hline
F090&HD 33564&12&11&179.&26&1.1&20.886$\pm$0.092&?&SI LE&21 Jan\\
F092&HD 3196&13&7&ind.&ind.&1.9&21.109$\pm$0.300&N&SC\\
F096&HD 739&8&8&ind.&ind.&1.0&21.281$\pm$0.122&N&SC\\
F097&HD 89449&5&2&ind.&52&2.5&21.372$\pm$0.110&N&SI PI TI BE&23 Jan.\\
F099&HD 160032&50&30&ind.&ind.&1.7&21.447$\pm$0.152&Y&MA\\
F101&HD 22001&7&8&ind.&ind.&0.9&21.681$\pm$0.056&Y&TI\\
F102&HD 16673&24&12&ind.&ind.&2.0&21.763$\pm$0.194&N&LE\\
F103&HD 108954&54&42&12&16&1.3&21.782$\pm$0.166 &N&SI&23 Jan\\
F106&HD 206826&6&11&ind.&ind.&0.5&22.204$\pm$0.211&N&LE\\
F108&HD 106516&4&12&ind.&ind.&0.3&22.336$\pm$0.404&N&LE\\
F109&HD 68146&0&30&ind.&ind.&0.0&22.377$\pm$0.150&N&MA\\
F111&HD 213845&9&8&ind.&ind.&1.1&22.681$\pm$0.134&N&SC\\
F112&HD 16765&21&12&ind.&ind.&1.8&22.687$\pm$0.428&N&SC\\
F113&HD 89125&12&13&151&11.&0.9&22.789$\pm$0.181&N&SI LE&23 Jan\\
F114&HD 168151&15&10&ind.&ind.&1.5&22.906$\pm$0.089&N&LE\\
F115&HD 162003&28&17&ind.&ind.&1.6&22.918$\pm$0.177&N&LE\\
F117&HD 219571&5&5&ind.&ind.&1.0&23.066$\pm$0.335&N&SC\\
F118&HD 160922&14&16&ind.&ind.&0.9&23.165$\pm$0.091&N&LE\\
F119&HD 11171&16&6&ind.&ind.&2.7&23.175$\pm$0.139 &Y&SC\\
F120&HD 101177&145&49&45.&10.&2.9&23.195$\pm$0.391&N&SI&22 Jan\\
F121&HD 100180&124&63&47.&17.&2.2&3.326$\pm$0.658&N&SI& 08 Dec\\
F122&HD 7439&5&7&ind.&ind.&0.7&23.375$\pm$0.164&Y&SC AP\\
F124&HD 4676&3&23&ind.&ind.&0.1&23.451$\pm$0.148&N&LE\\
F126&HD 214953&25&15&ind.&ind.&1.7&23.621$\pm$0.223&N&SC BM\\
A002&HD 187642&7&6&ind.&ind.&1.2&5.125$\pm$0.014&N&PI SC TI BE\\
A005&HD 102647&6&7&ind.&ind.&0.9&11.011$\pm$0.063&Y&TI PI BE BM\\
A006&HD 60179&67&25&179.&10.&2.7&14.005$\pm$0.408&N&SI&08 Dec\\
A007&HD 76644&24&11&93&28&2.2&14.509$\pm$0.034&N&SI LE&08 Dec\\
A011&HD 97603&7&7&175&20&1.0&17.918$\pm$0.080&N&SI PI TI BE&23 Jan\\
A012&HD 11636&12&5&ind.&ind.&2.4&17.965$\pm$0.187&N&PI TI BE\\
A013&HD 115892&11&7&ind.&ind.&1.6&18.021$\pm$0.055&N&TI BM\\
A015&HD 141795&16&12&ind.&ind.&1.3&21.610$\pm$0.089&N&LE\\
A016&HD 38678&11&8&131&26&1.4&21.612$\pm$0.075&Y&SI TI&08 Dec\\
A018&HD 139006&14&6&ind.&ind.&2.3&23.007$\pm$0.148&Y&PI BE+ BM\\
A019&HD 156164&18&29&ind.&ind.&0.6&23.038$\pm$0.080&N&BE\\
A021&HD 2262&10&5&ind.&ind.&2.0&23.807$\pm$0.091 &Y&TI MA\\
A023&HD 16970&5&14&ind.&ind.&0.4&24.348$\pm$0.367&N&LE\\
A024&HD 95418&13&6&ind.&ind.&2.2&24.455$\pm$0.096&Y&PI TI BE BM\\
A028&HD 116657&18&15&ind.&52&1.2&26.309$\pm$0.579&?&SI LE&02 Mar\\
&13240+5456 A&35&36&ind.&ind.&1.0&25.064$\pm$0.088&?&SI&02 Mar\\
A029&HD 99211&32&17&ind.&ind.&1.9&25.246$\pm$0.127&N&LE\\
A032&HD 103287&30&41&4.1&29.2&0.8&25.510$\pm$0.26&N&SI&22,23 Jan\\
A034&HD 165777&19&7&ind.&ind.&2.7&26.620$\pm$0.156&N&TI\\
A035&HD 108767&28&29&ind.&ind.&1.0&26.637$\pm$0.113&N&BE\\
A038&HD 18978&6&6&ind.&ind.&1.0&27.168$\pm$0.140&N&SC\\
A039&HD 180777&4&10&ind.&ind.&0.4&27.303$\pm$0.142&N&LE\\
A040&HD 33111&9&7&ind.&ind.&1.3&27.362$\pm$0.314&N&TI AP\\
A041&HD 210418&9&7&ind.&ind.&1.3&28.180$\pm$0.671&N&TI PI BE\\
A042&HD 87696&13&13&32&22&1.0&28.238$\pm$0.144&Y&SI PI BE&04 Dec\\
A045&HD 78209&107&77&119&19&1.4&28.818$\pm$0.208&N&SI&04 Dec\\
A048&HD 125161&24&28&ind.&52&0.9&29.067$\pm$0.161 &N&SI&02 Mar\\
A049&HD 50241&12&8&ind.&ind.&1.5&29.398$\pm$1.528&N&TI\\
A052&HD 159560&20&20&ind.&ind.&1.0&30.351$\pm$0.106&N&LE\\
A053&HD 125162&14&8&21&22&1.8&30.355$\pm$0.147&Y&SI LE&02 Mar\\
A056&HD 56537&50&31&39&12&1.6&30.888$\pm$0.210&N&SI&08 Dec\\
A063&HD 222603&21&13&ind.&ind.&1.6&32.681$\pm$0.203 &N&LE BE\\
A064&HD 20320&53&21&ind.&ind.&2.5&33.650$\pm$0.328 &Y&LE\\
A065\tnote{e}&15278+2906A&28&31&148&56&0.9&34.281$\pm$0.892 & N&SI&02 Mar \\
A066&HD 104513&25&42&ind.&52&0.6&34.282$\pm$0.881 &N&SI&22 Jan\\
A067&HD 14055&84&47&18&21&1.8&34.397$\pm$0.284 &Y&SI&08 Dec\\
A068&HD 91312&29&40&122&19&0.7&34.627$\pm$0.623&?&SI&04 Dec\\
\end{tabular}
\end{onecolumn}
\end{center}
\end{ThreePartTable}
\end{table*}

\begin{table*}		
\begin{ThreePartTable}
\contcaption{Polarization data for all the stars observed at OMM and from the Leroy compilation with a polarization $P < 3\sigma_P$.}
\begin{center}
\begin{onecolumn}
\begin{tabular}{clrrrcllcrl}
\hline
UNS  & name & $P$ & $\sigma_P$ & $\theta$(\degr) & $\sigma_\theta$(\degr) & 
$P/\sigma_P$ & Distance &FIR\tnote{a} &Observer\tnote{b}& Date\tnote{c}\\
ID &&($10^{-5}$)&($10^{-5}$)&&&&(pc)&excess&&\\
\hline\hline
A069&HD 112412&24&13&109&22&1.8&36.900$\pm$5.523&?&SI LE&02 Mar\\
&HD 112413&31&42&ind.&ind.&0.7&35.247$\pm$1.093&?&SI&02 Mar\\
A074&HD 79439&124&71&60&16&1.7&35.837$\pm$0.257&N&SI&08 Dec\\
K077&HD 214749&22&23&ind.&ind.&1.0&36.367$\pm$0.528&N&SC\\
A078&HD 184006&101&51&133&16&2.0&37.216$\pm$0.152&N&SI&02 Mar\\
A079&HD 102124&20&29&8&30&0.7&37.411$\pm$0.350&N&SI&02 Mar\\
A080&HD 177196&9&11&ind.&ind.&0.8&37.434$\pm$0.238&N&LE\\
A082&HD 71155&0&43&ind.&52&0.0&37.514$\pm$0.267&Y&SI&22 Jan\\
A083&HD 80081&7&4&ind.&52&1.8&38.181$\pm$1.119&N&SI PI BE HA&21 Jan\\
A084&HD 78045&23&8&ind.&ind.&2.9&38.283$\pm$0.176&N&TI\\
A086&HD 13161&46&24&ind.&52&1.9&38.865$\pm$0.514&Y&SI BE&23 Jan\\
A087&HD 95608&33&32&86&16&1.0&38.956$\pm$0.258&N&SI&23 Jan\\
A089&HD 215789&9&4&ind.&ind.&2.2&39.497$\pm$0.748&N&SC TI&\\
A090&HD 5448&34&12&ind.&ind.&2.8&39.602$\pm$1.341&N&TI BE\\
A101&HD 130109&50&30&ind.&ind.&1.7&41.244$\pm$0.323&N&MA BE\\
A103&HD 1404&18&29&ind.&ind.&0.6&41.291$\pm$0.358&Y&BE\\
A104&HD 90132&70&30&ind.&ind.&2.3&41.477$\pm$0.464&N&MA\\
A110&HD 89021&9&8&ind.&52&1.1&42.129$\pm$1.378&N&SI BE&22 Jan\\
A113&HD 23281&29&18&ind.&ind.&1.6&42.391$\pm$0.898&Y&LE\\
A118&HD 15008&9&5&ind.&ind.&1.8&42.809$\pm$0.623&N&SC\\
A123&HD 213398&14&6&ind.&ind.&2.3&43.804$\pm$0.422&Y&SC\\
A127&HD 140436&28&29&ind.&52&1.0&44.621$\pm$1.010 &N&SI BE&02 Mar\\
 &HR 8799&70&25&95.5&8.9&2.8&39.4 $\pm$0.1 &Y&SI& 08 Dec\\
\hline

\end{tabular}

\begin{tablenotes}
\item[a] Detection of FIR excess: yes (Y), no (N), no information or uncertain (?).
  \item[b] The key for the observers is the same as in Table~\ref{tab:3sigma}.
  \item[c] Observations from OMM were obtained during the winter 2009-2010.
  \item[d] When the uncertainty on the polarization angle is larger than $\approx 52\degr$, its orientation is indefinite (see section 3). Also it is customary not to give the polarization angle when the polarization is considered to be too small to yield a reliable polarization angle. 
  \item[e] See the note about A065 in Table~\ref{tab:3sigma}.
\end{tablenotes}
\end{onecolumn}
\end{center}

\end{ThreePartTable}
\end{table*}

\label{lastpage}

\end{document}